\documentclass[12pt]{article}
\pdfoutput=1
\usepackage{jheppub}
\usepackage{mathtools,amssymb,amsthm}
\usepackage[utf8]{inputenc}
\usepackage{enumitem}
\usepackage{dsfont}

\begin{document}

\newcommand{\be}{\begin{equation}}
\newcommand{\ee}{\end{equation}}
\newcommand{\bi}{\begin{itemize}}
\newcommand{\ei}{\end{itemize}}
\def\ba#1\ea{\begin{align}#1\end{align}}
\def\bg#1\eg{\begin{gather}#1\end{gather}}
\def\bm#1\em{\begin{multline}#1\end{multline}}
\def\bmd#1\emd{\begin{multlined}#1\end{multlined}}

\def\a{\alpha}
\def\b{\beta}
\def\c{\chi}
\def\C{\Chi}
\def\d{\delta}
\def\D{\Delta}
\def\e{\epsilon}
\def\ve{\varepsilon}
\def\g{\gamma}
\def\G{\Gamma}
\def\h{\eta}
\def\k{\kappa}
\def\l{\lambda}
\def\L{\Lambda}
\def\m{\mu}
\def\n{\nu}
\def\p{\phi}
\def\P{\Phi}
\def\vp{\varphi}
\def\q{\theta}
\def\Q{\Theta}
\def\r{\rho}
\def\s{\sigma}
\def\S{\Sigma}
\def\t{\tau}
\def\u{\upsilon}
\def\U{\Upsilon}
\def\w{\omega}
\def\W{\Omega}
\def\x{\xi}
\def\X{\Xi}
\def\y{\psi}
\def\Y{\Psi}
\def\z{\zeta}
\def\cL{{\mathcal L}}
\newcommand{\vn}{{\vec n}}

\newcommand{\la}{\label}
\newcommand{\ci}{\cite}
\newcommand{\re}{\ref}
\newcommand{\er}{\eqref}
\newcommand{\se}{\section}
\newcommand{\sse}{\subsection}
\newcommand{\ssse}{\subsubsection}
\newcommand{\fr}{\frac}
\newcommand{\na}{\nabla}
\newcommand{\pa}{\partial}
\newcommand{\td}{\tilde}
\newcommand{\wtd}{\widetilde}
\newcommand{\ph}{\phantom}
\newcommand{\eq}{\equiv}
\newcommand{\wg}{\wedge}
\newcommand{\cd}{\cdots}
\newcommand{\nn}{\nonumber}
\newcommand{\qu}{\quad}
\newcommand{\qqu}{\qquad}
\newcommand{\lt}{\left}
\newcommand{\rt}{\right}
\newcommand{\lra}{\leftrightarrow}
\newcommand{\ol}{\overline}
\newcommand{\ap}{\approx}
\renewcommand{\(}{\left(}
\renewcommand{\)}{\right)}
\renewcommand{\[}{\left[}
\renewcommand{\]}{\right]}
\newcommand{\<}{\langle}
\renewcommand{\>}{\rangle}
\newcommand{\Hc}{\mathcal{H}_{code}}
\newcommand{\HR}{\mathcal{H}_R}
\newcommand{\HRb}{\mathcal{H}_{\ol{R}}}
\newcommand{\lan}{\langle}
\newcommand{\ran}{\rangle}
\newcommand{\Hra}{\mathcal{H}_{W_\alpha}}
\newcommand{\Hrba}{\mathcal{H}_{\ol{W}_\alpha}}

\newcommand{\tr}{\operatorname{tr}}
\newcommand{\Tr}{\operatorname{Tr}}
\newcommand{\bH}{{\mathbb H}}
\newcommand{\bR}{{\mathbb R}}
\newcommand{\bZ}{{\mathbb Z}}
\newcommand{\cA}{{\mathcal A}}
\newcommand{\cB}{{\mathcal B}}
\newcommand{\cC}{{\mathcal C}}
\newcommand{\cE}{{\mathcal E}}
\newcommand{\cI}{{\mathcal I}}
\newcommand{\cN}{{\mathcal N}}
\newcommand{\cO}{{\mathcal O}}
\newcommand{\zb}{{\bar z}}
\newcommand{\LL}{\mathcal{L}}

\newcommand{\Area}{\operatorname{Area}}
\newcommand{\ext}{\operatorname*{ext}}
\newcommand{\total}{\text{total}}
\newcommand{\bulk}{\text{bulk}}
\newcommand{\brane}{\text{brane}}
\newcommand{\matter}{\text{matter}}
\newcommand{\Wald}{\text{Wald}}
\newcommand{\anomaly}{\text{anomaly}}
\newcommand{\extrinsic}{\text{extrinsic}}
\newcommand{\gen}{\text{gen}}
\newcommand{\mc}{\text{mc}}
\renewcommand{\th}{\text{th}}

\newcommand{\T}[3]{{#1^{#2}_{\ph{#2}#3}}}
\newcommand{\Tu}[3]{{#1_{#2}^{\ph{#2}#3}}}
\newcommand{\Tud}[4]{{#1^{\ph{#2}#3}_{#2\ph{#3}#4}}}
\newcommand{\Tdu}[4]{{#1_{\ph{#2}#3}^{#2\ph{#3}#4}}}
\newcommand{\Tdud}[5]{{#1_{#2\ph{#3}#4}^{\ph{#2}#3\ph{#4}#5}}}
\newcommand{\Tudu}[5]{{#1^{#2\ph{#3}#4}_{\ph{#2}#3\ph{#4}#5}}}

\title{
One-loop universality of holographic codes
}
\author{Xi Dong}
\author{and Donald Marolf}
\affiliation{Department of Physics, University of California, Santa Barbara, CA 93106, USA}
\emailAdd{xidong@ucsb.edu}
\emailAdd{marolf@ucsb.edu}

\abstract{
Recent work showed holographic error correcting codes to have simple universal features at $O(1/G)$.  In particular, states of fixed Ryu-Takayanagi (RT) area in such codes are associated with flat entanglement spectra indicating maximal entanglement between appropriate subspaces.  We extend such results to one-loop order ($O(1)$ corrections) by controlling both higher-derivative corrections to the bulk effective action and dynamical quantum fluctuations below the cutoff.  This result clarifies the relation between the bulk path integral and the quantum code, and implies that i) simple tensor network models of holography continue to match the behavior of holographic CFTs beyond leading order in $G$, ii) the relation between bulk and boundary modular Hamiltonians derived by Jafferis, Lewkowycz, Maldacena, and Suh holds as an operator equation on the code subspace and not just in code-subspace expectation values, and iii) the code subspace is invariant under an appropriate notion of modular flow.  A final corollary requires interesting cancelations to occur in the bulk renormalization-group flow of holographic quantum codes. Intermediate technical results include showing the Lewkowycz-Maldacena computation of RT entropy to take the form of a Hamilton-Jacobi variation of the action with respect to boundary conditions, corresponding results for higher-derivative actions, and generalizations to allow RT surfaces with finite conical angles.
}
\maketitle

\section{Introduction}

There has been much recent interest in the idea that the bulk/boundary dictionary of AdS/CFT represents a quantum error correcting (QEC) code  \cite{Almheiri:2014lwa,Pastawski:2015qua,Hayden:2016cfa,Dong:2016eik,Harlow:2016vwg,Cotler:2017erl,Bao:2017guc,Hayden:2018khn}.  According to this paradigm, full recovery of standard bulk physics can occur only on a `code subspace' ${\cal H}_{\rm code}$ in the CFT Hilbert space ${\cal H}_{\rm CFT}$.   Consistent with either the firewall \cite{Almheiri:2012rt,Almheiri:2013hfa,Marolf:2013dba,Marolf:2015dia} or state-dependent observables \cite{Papadodimas:2012aq,Verlinde:2012cy,Papadodimas:2013jku,Verlinde:2013qya,Verlinde:2013uja,Papadodimas:2015jra} hypotheses, the orthogonal complement of ${\cal H}_{\rm code}$ is presumed to contain states describing generic black holes inside which at least any given code will fail to reconstruct  standard bulk physics.

The arguments \cite{Dong:2016eik,Harlow:2016vwg} for this paradigm are strong when one considers the more restricted subspaces ${\cal H}_\phi \subset {\cal H}_{\rm CFT}$ whose bulk duals describe small quantum fluctuations at lowest non-trivial order in the bulk Newton constant $G$ around a given classical solution $\phi$.  In that context, a QEC structure with code subspace ${\cal H}_\phi$ follows from the one-loop Faulkner-Lewkowycz-Maldacena relation \cite{Faulkner:2013ana}.  In particular, given any partition of a CFT Cauchy surface into regions $R$ and $\ol R$, one obtains a code with a property known as complementary recovery.  We will review this structure in section \ref{sec:code} below.

It is natural to expect that such codes can be sewn together into a single code on ${\cal H}_{\rm code} := \oplus_\phi
{\cal H}_\phi$.  While the details of this operation remain to be understood, the recent works \cite{Akers:2018fow,Dong:2018seb} discovered strong similarities between these codes at leading order in $G$.  The point here is that codes with complementary recovery are characterized by their pattern of entanglement between appropriate factors of ${\cal H}_{\rm CFT}$, and  at leading order in $G$ Ref.~\cite{Akers:2018fow,Dong:2018seb} showed this pattern to be the same in each ${\cal H}_\phi$ up to unitary transformations.  Specifically, at this order states of definite area for the relevant Ryu-Takayanagi (RT) surface \cite{Ryu:2006bv,Ryu:2006ef} -- or more generally the Hubeny-Rangamani-Takayanagi (HRT) surface \cite{Hubeny:2007xt} --  always induce density matrices on each tensor factor that are proportional to projection operators.  In other words, referring to this leading order as $O(1/G)$ one may say that at $O(1/G)$ such density matrices take a universal form with a ``flat'' spectrum of eigenvalues $\lambda_k$, meaning that the $\lambda_k$ are independent of $k$ for $\lambda_k \neq 0$.  Note that one may equivalently say that at $O(1/G)$ every such code involves  maximal entanglement between subspaces of the tensor factors, or alternatively that the associated density matrices are proportional to projection operators.

Our present work refines this result by establishing a sense in which it extends to $O(1)$.  Since we treat bulk gravity as an effective field theory with a cut-off, there are two different $O(1)$ effects to consider.  The first comes from higher derivative corrections to the bulk effective action at some cut-off scale.  Such corrections contain effects of ultraviolet (UV) quantum fluctuations at energies above the cut-off that have been integrated out. As is well known, such higher derivative terms in the action cause the geometric entropy associated with the bulk entangling (RT or HRT) surface to differ from $A/4G$ by related higher derivative terms \cite{Dong:2013qoa,Camps:2013zua,Miao:2014nxa,Dong:2017xht}. The second $O(1)$ effect comes from infrared (IR) bulk quantum fluctuations at energies below the cut-off which remain to be integrated over in the path integral.

Though the two effects are physically related, they enter the code formalism in qualitatively different ways.  Indeed, as we discuss in section \ref{sec:code}, it is natural to conjecture that dynamical IR quantum fluctuations merely determine which state in the code subspace arises from a given path integral, and thus that such fluctuations may be completely ignored when computing certain entanglement properties of the code itself. In effect, for such purposes one would then treat the effective action defined at the cutoff scale as a classical variational principle.  Our arguments below will verify that this conjecture is correct.

However, it will first be necessary to deal with the higher derivative corrections to the effective action at the cutoff scale.  This is done in section \ref{sec:HD} (with help from appendices \ref{app:LMVPRT} and \ref{app:HDI}) by first reformulating the Lewkowycz-Maldacena procedure \cite{Lewkowycz:2013nqa} for computing Ryu-Takayanagi gravitational entanglement for two-derivative Einstein-Hilbert gravity.  Indeed, we show that their computation can be interpreted as a Hamilton-Jacobi variation of an action with respect to boundary conditions, where in this case the role of the boundary condition is played by a choice of conical defect angle $\delta$ on the RT surface, and that the gravitational entanglement remains $A/4G$ even on Euclidean saddles with $\delta \neq 0$.  In particular, this confirms that the Lewkowycz-Maldacena procedure is a direct generalization of the Carlip-Teitelboim approach to black hole entropy \cite{Carlip:1993sa} to cases that break the $U(1)$ symmetry of \cite{Carlip:1993sa}. Further extending this result to arbitrary higher derivative actions allows one to repeat the arguments of \cite{Dong:2018seb} and show that treating the effective action as a classical variational principle would again yield density matrices proportional to projectors for states of fixed geometric entropy $\sigma = A/4G + \text{(higher derivative corrections)}$.

It then remains to properly address the dynamical IR quantum fluctuations.  We do so in section \ref{sec:result} by considering states $|\psi \rangle_\s$ of fixed geometric entropy and tracing them over $\ol R$ to define density matrices $\rho_R$.  Taking the tensor product with the identity operator ${\mathds 1}_{\ol R}$ on $\ol R$ yields an operator $\rho_R \otimes {\mathds 1}_{\ol R}$ on ${\cal H}_{\rm CFT}$.  Using the results from section \ref{sec:HD}, for $|\psi\rangle_\s \in {\cal H}_\phi$ we show $\rho_R \otimes {\mathds 1}_{\ol R}$ to
preserve an appropriately defined ${\cal H}_\phi$.  It then follows immediately that density matrices on $R$ defined by the code itself must again be proportional to projection operators.  The universal flat entanglement spectrum \textit{of the code itself} is thus maintained at one-loop order, even though generic encoded states no longer have flat entanglement.  A corollary is confirmation of the above-mentioned conjecture that dynamical IR quantum fluctuations merely determine which state in the code subspace arises from a given path integral and that properties of the code itself are determined by treating the cutoff-scale effective action as a classical variational principle.

We conclude in section \ref{sec:disc} with discussion focusing on implications for the renormalization group (RG) flow of holographic quantum codes and for the relation between bulk and boundary modular Hamiltonians derived by Jafferis, Lewkowycz, Maldacena, and Suh (JLMS) \cite{Jafferis:2015del}. In the former context, our result implies precise cancelations between a number of different effects.  In the latter context, it shows that their relation holds as an operator statement on each ${\cal H}_\phi$ and not just in code-subspace expectation values (see discussion in \cite{Dong:2016eik} and \cite{Dong:2018seb}).  Although the structure of QEC with complementary recovery is expected to break down beyond one-loop order, many of our arguments nevertheless remain valid more generally and must thus constrain any structure that remains.

\section{Review of holographic quantum codes}
\label{sec:code}

It is useful to briefly review the role of quantum error correcting codes in holography.  Our discussion largely follows that of \cite{Dong:2018seb}, which is in turn based on \cite{Almheiri:2014lwa,Dong:2016eik,Harlow:2016vwg}.  However,
since higher derivative corrections play a key role in the remainder of this work, we take care to emphasize the relation to bulk effective field theory concepts and, in particular, the evolution of the code under bulk renormalization-group flow.  Such issues were also mentioned in \cite{Faulkner:2013ana,Akers:2018fow,Dong:2018seb}, but we wish to place them front and center.

As in the introduction, we focus on subspaces ${\cal H}_\phi \subset {\cal H}_{\rm CFT}$ whose bulk duals describe small quantum fluctuations at lowest non-trivial order in the bulk Newton constant $G$ around a given classical solution $\phi$.  Given any partition of a CFT Cauchy surface into regions $R$ and $\ol R$ and the associated RT/HRT-surface $\gamma_R$ in the bulk spacetime, one may define the corresponding bulk entanglement wedges $W$ (${\ol W}$)
\cite{Czech:2012bh,Wall:2012uf,Headrick:2014cta}
as the bulk domain of dependence of any achronal bulk surface whose boundary is $R \cup \gamma_R$ ($\ol R \cup \gamma_R$).
States in ${\cal H}_\phi$ then must obey a Faulkner-Lewkowycz-Maldacena (FLM) relation
\be \label{FLM}
S({\rho}_R)=\Tr\left(\rho_{W} \LL_R\right)+S_{W}(\rho_{W}),
\ee
where $\rho_R$ is the CFT density matrix obtained by tracing over $\ol R$ and $\rho_{W}$ is the density matrix describing bulk quantum fields in $W$.  The operator $\LL_R$ is localized on the RT/HRT-surface  $\gamma_R$ and takes the form $A[\gamma_R]/(4G) + \dots$ where $\dots$ represents appropriate higher derivative corrections.  The entropy $S({\rho}_R)$ is computed as usual in the CFT, but $S_{W}(\rho_{W})$ is the entropy of the bulk state $\rho_{W}$ defined as a linear functional on a von Neumann algebra $M_W$ of operators in $W$. The operator $\LL_R$ is an element of $M_W$, and by interchanging $R$ and $\ol R$ it turns out also an element of the algebra $M_{\ol W}$ on ${\ol W}$.  As a result, $\LL_R$ commutes with all operators in $M_W$ and thus lies in the center of $M_W$.  Note that $M_{\ol W}$ is the commutant of $M_W$ in the algebra of bulk fields, and that $M_W$ is also the commutant of $M_{\ol W}$.

A key point for our work below is that Ref.~\cite{Faulkner:2013ana} derived \eqref{FLM} using the bulk path integral, and in particular treated the bulk as an effective field theory.  One should thus understand \cite{Faulkner:2013ana} to rely on having a bulk effective action valid at locally-measured energies below some bulk cutoff scale $\Lambda$.  In particular, the operator $\LL_R$ is determined by applying the Lewkowycz-Maldacena procedure \cite{Lewkowycz:2013nqa} to this effective action and so also depends on $\Lambda$; see \cite{Dong:2013qoa,Camps:2013zua,Miao:2014nxa,Dong:2017xht} for treatments of higher derivative corrections.  We will discuss this procedure in more detail in section \ref{sec:HD}, but for now we note that, although dynamical fluctuations below the cutoff $\Lambda$ contribute to the expectation value of $\LL_R$ in \eqref{FLM},  the procedure determining the {\it form} of $\LL_R$ is entirely classical and makes no reference to these fluctuations.  As a result, the expression for $\LL_R$ is precisely given by the {\it classical} geometric entropy defined by the effective action at the scale $\Lambda$.

Due to our high energy cutoff, we assume that we can treat our bulk theory in parallel with quantum mechanics on a finite-dimensional Hilbert space -- perhaps by imposing further cutoffs as well.  In that context it follows that any action of a von Neumann algebra $M_W$ on a Hilbert space $\mathcal{H}_\phi$ allows one to decompose $\mathcal{H}_\phi$ as
\be\label{eq:hilbertdecomp}
\mathcal{H}_\phi=\oplus_{\alpha\in S}\left(\mathcal{H}_{W_\alpha}\otimes \mathcal{H}_{\ol{W}_\alpha}\right),
\ee
where the decomposition defines $\mathcal{H}_{W_\alpha}$ and $\mathcal{H}_{\ol{W}_\alpha}$, $S$ is an appropriate index set, and operators in either $M_W$ or its commutant $M_{\ol W}$ are block diagonal in $\alpha$; see e.g.\ the appendix of \cite{Harlow:2016vwg}. See also related comments in \cite{Giddings:2019hjc}. We may also choose the tensor factorization within each block such that $M_W$ ($M_{\ol W}$) contains precisely those operators that act trivially on $\Hrba$  ($\Hra$).  The intersection $Z = M_W \cap M_{\ol W}$ gives the center of both $M_W$ and $M_{\ol W}$ and contains block diagonal matrices that are proportional to the identity on $\Hra\otimes\Hrba$ within each block\footnote{\label{foot:Hphi}In fact, as we discuss in section \ref{sec:result} below, equation \eqref{eq:hilbertdecomp} has some tension with the context just discussed.  In particular, the right-hand side contains exact eigenstates of $\alpha$, but such eigenstates cannot be described as small quantum fluctuations around a classical background $\phi$.  We will resolve this tension in section \ref{sec:result} by slightly generalizing the definition of $\mathcal{H}_\p$ so that it contains such $\a$-eigenstates, noting that the derivation of \eqref{FLM} holds equally well on these states.}.  We may thus write
\begin{align}
\label{eq:VNalgebras}\nonumber
M_W&=\oplus_\alpha\left(\LL(\Hra)\otimes {\mathds 1}_{\ol{W}_\alpha}\right),\\
M_{\ol W}&=\oplus_\alpha\left({\mathds 1}_{W_\alpha}\otimes \LL(\Hrba)\right),\\\nonumber
Z&=\oplus_\alpha \LL({\mathds C}) {\mathds 1}_{W_\alpha \ol{W}_\alpha},
\end{align}
where $\LL(\mathcal{H})$ denotes the set of linear operators on the Hilbert space $\mathcal{H}$.  We refer to $\alpha$ as the superselection parameter below.

Since the above structure follows from \eqref{FLM}, it is again valid only below some cutoff $\Lambda$.  While $\Lambda$ is to some extent arbitrary, we should expect the Hilbert spaces, the decomposition \eqref{eq:hilbertdecomp}, and the algebras of operators to depend the value of $\Lambda$ that is chosen.  This is especially true for the operator $\LL_R$, whose form depends on the effective action as noted above and which -- as with all operators in $M_W, M_{\ol W}$ -- should be thought of as being smeared over length scales $1/\Lambda$.  In particular, all of these structures can experience non-trivial renormalization-group flows under changes in $\Lambda$.

As a further comment on \eqref{eq:VNalgebras}, we note that if the bulk were described by a scalar field theory, we could choose the algebra $M_W$ so that the center $Z$ is trivial, containing only operators proportional to the identity on $\mathcal{H}_\phi$.  The index set $S$ would then contain only one element so that \eqref{eq:hilbertdecomp} becomes a simple tensor product.  But the bulk is described by a theory of gravity, and the resulting diffeomorphism gauge symmetry implies constraints that forbid  quantum states (or even classical initial data) in $W$ and ${\ol W}$ from being chosen independently.  In this context the set $S$ is generally non-trivial and -- as in the case of Yang-Mills theories --
taking $M_W$ to be the algebra of gauge-invariant operators in $W$ yields a non-trivial center
\cite{Donnelly:2011hn,Casini:2013rba,Donnelly:2014gva,Donnelly:2015hxa,Harlow:2016vwg}.

In contrast, we will ignore issues associated with constraints in the CFT dual and write
\begin{equation}
{\cal H}_{\rm CFT} = {\cal H}_R \otimes {\cal H}_{\ol R}.
\end{equation}
While the dual CFT is often a gauge theory and thus {\it does} have similar issues involving constraints and a lack of factorization, any such CFT gauge symmetry is expected to be unrelated to bulk diffeomorphism invariance.  As a result, the corresponding central operators in the CFT will not directly relate to the bulk center $Z$ discussed here. Following standard practice, we thus ignore this complication in the present discussion.

Returning to the bulk,
we can now explain the entropy $S_{W}(\rho_{W})$ in more detail.  Since $\alpha$ denotes the eigenvalues of center operators, the density matrix $\rho_{W}$ must take the block-diagonal form
\begin{equation}
\rho_{W} = \oplus_\alpha p_\alpha \rho_{W_\alpha},
\end{equation}
where $\Tr \rho_{W_\alpha}=1$ and $\sum_\alpha p_\alpha=1$.
The desired entropy is then simply
\be
S_{W}(\rho_{W}) =  -\sum_\alpha p_\alpha \log p_\alpha +\sum_\alpha p_\alpha S(\rho_{W_\alpha}).\label{algent}
\ee

As shown in \cite{Harlow:2016vwg}, the FLM formula \eqref{FLM} tightly constrains the relations between the bulk factors $\mathcal{H}_{W_\alpha},\mathcal{H}_{\ol{W}_\alpha}$ and the CFT factors
${\cal H}_R, {\cal H}_{\ol R}$.  In particular, if all states in a
 code subspace $\mathcal{H}_\phi \subset \mathcal{H}_R \otimes \mathcal{H}_{\ol{R}}$ satisfy
\eqref{FLM} and its analogue for $\ol R$,
then  $\HR$ and $\HRb$ must admit decompositions of the form
\begin{align}\nonumber
\HR&=\oplus_\alpha\left(\mathcal{H}_{R_\alpha^1}\otimes \mathcal{H}_{R_\alpha^2}\right)\oplus \mathcal{H}_{R_3},\\
\la{hdecomp}
\HRb&=\oplus_\alpha\left(\mathcal{H}_{\ol{R}_\alpha^1}\otimes \mathcal{H}_{\ol{R}_\alpha^2}\right)\oplus \mathcal{H}_{\ol{R}_3},
\end{align}
where $\mathcal{H}_{R_\alpha^1}\cong \Hra$ and $\mathcal{H}_{\ol{R}_\alpha^1}\cong \Hrba$ with $\cong$ denoting Hilbert space isomorphisms. Furthermore,
one can choose a  basis $|{\alpha, i j}\ran$ of $\mathcal{H}_\phi$ associated with the decomposition \eqref{eq:hilbertdecomp}.  In particular we may take
\be\label{basisrep}
|{\alpha, ij}\ran=U_R U_{\ol{R}}\left(|\alpha,i\ran_{R_\alpha^1}\otimes |\alpha,j\ran_{\ol{R}_\alpha^1}\otimes |\chi_\alpha\ran_{R^2_\alpha \ol{R}_\alpha^2}\right)
\ee
for some unitaries $U_R$, $U_{\ol{R}}$ on $\HR$ and $\HRb$, bases $\{|\alpha,i\rangle\}, \{|\alpha, j\rangle\}$ of $\mathcal{H}_{R_\alpha^1}, \mathcal{H}_{\ol{R}_\alpha^1}$, and some set of states $|\chi_\alpha\ran \in \mathcal{H}_{R_\alpha^2} \otimes \mathcal{H}_{\ol{R}_\alpha^2}$.  Ref.  \cite{Harlow:2016vwg} called such codes ``operator algebra quantum error-correcting codes with complementary recovery'', as \eqref{basisrep} is equivalent to the requirement that the action of any operator in $M_W$ on a state in ${\cal H}_\phi$ can be reproduced by acting on that state with an operator in $R$, and correspondingly for $M_{\ol W}$, $\ol R$.

Since arbitrary unitaries on $\mathcal{H}_{R_\alpha^1}$, $\mathcal{H}_{\ol{R}_\alpha^1}$, ${\cal H}_{R_\alpha^2}$, ${\cal H}_{\ol R_\alpha^2}$ can be absorbed into $U_R, U_{\ol R}$, the only independent structure in \eqref{basisrep} comes from the coefficients in the Schmidt decomposition of $|\chi_\alpha\ran_{R^2_\alpha \ol{R}_\alpha^2}$, or equivalently the spectrum of eigenvalues of the density matrix $\chi_{R_\alpha^2}\equiv \Tr_{\ol{R}_\alpha^2}|\chi_\alpha\ran\lan\chi_\alpha|$ (which is also the spectrum of $\chi_{\ol R_\alpha^2}\equiv \Tr_{R_\alpha^2}|\chi_\alpha\ran\lan\chi_\alpha|$).  This spectrum is thus the essence of any code, and it is this spectrum that was shown in \cite{Dong:2018seb} to be flat at $O(1/G)$ in states of fixed RT-area; see also \cite{Akers:2018fow}.

Tracing \eqref{basisrep} over $\ol R$ yields
\be\label{rhorep}
{\rho}_R=\sum_\alpha p_\alpha U_R\left(\rho_{R_\alpha^1}\otimes \chi_{R_\alpha^2}\right)U_R^\dagger,
\ee
where $\rho_{R_\alpha^1}$ is the image of $\rho_{W_\alpha}$ under the isomorphism between $\Hra$ and $\mathcal{H}_{R_\alpha^1}$.  Using \eqref{algent}, the von Neumann entropy of \eqref{rhorep} immediately takes the form  \eqref{FLM} with the identification
\be
\label{eq:idL}
\LL_R=\sum_\alpha S(\chi_{R_{\alpha}^2}){\mathds 1}_{W_\alpha \ol{W}_\alpha}.
\ee
As described in \cite{Dong:2018seb,Akers:2018fow} the entropies of the normalized density matrices $\rho_R^n/(\Tr{\rho_R^n})$ take a similar form, though they will satisfy \eqref{FLM} with the same identification \eqref{eq:idL} if and only if each $\chi_{R_\alpha^2}$ satisfies $\chi_{R_\alpha^2}^2 \propto \chi_{R_\alpha^2}$; i.e., if each such density matrix is proportional to a projection operator.

Note that eigenstates of the superselection parameter $\alpha$ are also eigenstates of $\LL_R$.  In holography, it is an interesting question whether $\alpha$ is defined completely by the eigenvalue of $\LL_R$ or whether it contains additional information,  but in either case let us simply consider an eigenstate of $\alpha$.  In such a state, if we for the moment ignore information within a distance $1/\Lambda$ (set by the cutoff scale $\Lambda$) away from the bulk entangling surface $\gamma_R$, the remaining information about bulk quantum fluctuations below the cutoff $\Lambda$ in a state $|\psi \rangle$ appears to be captured by the amplitudes $\langle \psi | \alpha, ij\rangle$ and the details of the state factors $|\alpha,i\ran_{R_\alpha^1}\otimes |\alpha,j\ran_{\ol{R}_\alpha^1}$; indeed, these ingredients suffice to determine the correlation functions of operators in $W\cup {\ol W}$.  The state $|\chi_\alpha \rangle_{R_\alpha^2 \ol R_\alpha^2}$ must thus be associated with bulk degrees of freedom with energies {\it above} the cutoff $\Lambda$.  Together with the sources at the AdS boundary, such high energy degrees of freedom determine a natural classical background on which dynamical quantum fluctuations propagate through the condition that the background be a stationary point of the effective action that arises from integrating them out\footnote{Note that the existence of a preferred classical background determined by a variational principle does not necessarily imply that quantum fluctuations around this background are small.}.  It is thus natural to conjecture that many properties of $|\chi_\alpha \rangle_{R_\alpha^2 \ol R_\alpha^2}$ can be computed by using the cutoff-scale effective action as a corresponding classical variational principle.  Indeed, the identification \eqref{eq:idL} shows that its von Neumann entanglement entropy can be calculated using the classical Lewkowycz-Maldacena procedure.  It is thus reasonable to expect this to extend to Renyi entropies $S_n(\chi_{R_\alpha^2})$.  Below, we refer to this idea as the classical effective action conjecture for the quantum code\footnote{We emphasize that our main results (derived in section \re{sec:result}) do not rely on this conjecture.  Rather, the conjecture is supported by -- and provides an intuitive way of understanding -- our main results.}.

We give a definitive, though somewhat indirect, argument for this result in section \ref{sec:result} below.  The rough sketch of the idea is to consider a state $|\psi\rangle_\sigma$ in an appropriate code subspace ${\cal H}_\phi$ such that $|\psi\rangle_\sigma$ is an eigenstate of ${\cal L}_R$ with eigenvalue $\sigma$.  We use this state to build a new state
\begin{equation}
\label{eq:cube}
|\psi_3\rangle_\sigma := \left(\rho_{R} \otimes {\mathds 1}_{\ol{R}} \right) |\psi\rangle_\sigma,
\end{equation}
where $\rho_{R}$ is the density matrix defined on $R$ by tracing $|\psi\rangle_\sigma$ over $\ol R.$  The new state is labelled with a subscript $3$ because ${\Tr}_{\ol R} \ |\psi_3\>_\sigma {}_\sigma\<\psi_3| = (\rho_R)^3$ and thus
\begin{equation}
\label{eq:cubenorm1}
{}_\sigma\<\psi_3|\psi_3\>_\sigma = {\Tr} \left(\rho_R^3\right).
\end{equation}

We use a bulk calculation to argue that $|\psi_3\rangle_\sigma$ also lies in the same code subspace ${\cal H}_\phi$.  Thus both states define the same density matrix
$\chi_{R_\alpha^2}\equiv \Tr_{\ol{R}_\alpha^2}|\chi_\alpha\ran\lan\chi_\alpha|$ on
${\cal H}_{R_\alpha^2}$ for each superselection sector $\alpha$ consistent with the fixed geometric entropy $\sigma$ (and on which $|\psi\rangle_\sigma$ has non-zero projection). The relation \eqref{eq:cube} then requires
 \begin{equation}
 \label{eq:chicube}
 \left(\chi_{R_\alpha^2}\right)^3 \propto \chi_{R_\alpha^2}.
\end{equation}
Since eigenvalues of density matrices are real and non-negative, the relation \eqref{eq:chicube} allows $\chi_{R_\alpha^2}$ to have only the eigenvalues $0$ and $1$ up to an overall normalization. Thus the density matrix $\chi_{R_\alpha^2}$ is proportional to a projector onto a subspace of dimension dictated by its entropy, which is in turn dictated by the associated eigenvalue of ${\cal L}_R$.  Using the decomposition \eqref{basisrep}, one can then more generally show that multiplication by $\left(\rho_{R} \otimes {\mathds 1}_{\ol{R}} \right)$ preserves the given code subspace ${\cal H}_\phi$.

We note that this result provides evidence supporting the above-mentioned classical effective action conjecture.  In particular, the bulk calculation deriving \eqref{eq:chicube} relies on properties of variational principles for higher-derivative actions that we will establish in section \ref{sec:HD} below.  These properties imply that with fixed geometric entropy, a purely classical (saddle-point) calculation of Renyi entropies would again give a flat entanglement spectrum for the $|\chi_\alpha \rangle_{R_\alpha^2 \ol R_\alpha^2}$ state, and thus that the associated density matrix on $R_\a^2$ would be a projector onto a subspace of dimension set by the associated saddle-point von Neumann entropy (i.e., by the geometric entropy). This is thus the prediction of the classical effective action conjecture, and we see that it agrees precisely with the results for the spectrum of $|\chi_\alpha \rangle_{R_\alpha^2 \ol R_\alpha^2}$ described above. 

\section{Higher derivative saddle-point Renyi entropies and states of fixed geometric entropy}
\label{sec:HD}

Before proceeding to the main argument in section \ref{sec:result}, we must first develop certain techniques for studying higher derivative actions and fixing the associated geometric entropy.  In particular, as mentioned above we will first need to show that a purely classical saddle-point treatment (ignoring dynamical quantum fluctuations) would give flat entanglement spectra.  As motivation for this result, recall from \cite{Akers:2018fow,Dong:2018seb}  that tracing states of fixed RT-area over $\ol R$ defines density matrices $\rho_R$ whose Renyi entropies
\begin{equation}
\label{eq:Renyi}
S_n(\rho_R) \equiv - \frac{1}{n-1} \log {\Tr} (\rho_R^n)
\end{equation}
are independent of $n$ at $O(1/G)$.  Were these Renyi entropies exactly constant, one could readily show all nonzero eigenvalues $\lambda_k$ of $\rho_R$ to be degenerate ($\lambda_k$ independent of $k$ for $\lambda_k \neq 0$).   Because the results of \cite{Akers:2018fow,Dong:2018seb} involved only the leading order behavior in $G$, it sufficed to consider saddle points of the Euclidean action.  Fluctuations about such saddles can contribute only at higher orders. In addition, the analysis of \cite{Akers:2018fow,Dong:2018seb} was limited to leading order in the inverse string tension $\alpha'$ as the bulk was assumed to be described by Einstein-Hilbert gravity with minimally-coupled matter fields.

Our purpose here is to extend such arguments to incorporate general higher derivative terms, including those representing higher order corrections in either $\alpha'$ or $G$.  In particular, in this section we again consider only contributions from the classical saddle-points themselves.   Discussion of possible contributions associated with fluctuations about such saddles will be deferred to section \ref{sec:result}.  We thus refer to the quantities computed below as saddle-point Renyi entropies $S_n^{\rm saddle}$.
As noted in section \ref{sec:code}, if we also fix the higher-derivative corrected geometric entropy $\sigma = A/4G + \dots$ to some value $\hat \sigma$,  it is natural to conjecture that $S_n^{\rm saddle}$ computes entropies of the code state $|\chi_\alpha \rangle \in {\cal H}_{R^2_\alpha} \otimes {\cal H}_{\ol R^2_\alpha}$ associated with the corresponding superselection sector $\alpha$.  We will argue that this is indeed the case in section \ref{sec:result} below.   Recall that $\sigma$ is specified by the choice of superselection sector $\alpha$, though we have left open the issue of whether $\alpha$ is fully specified by $\sigma$.

We will consider general higher derivative corrections which may involve an arbitrarily large number of derivatives in the effective action, for the following reasons.  In addition to large numbers of derivatives that can appear at high orders in $\alpha'$, it is important to note that moderately large numbers of derivative can appear at leading order in $\alpha'$ already in the one-loop corrections.  Indeed,
in bulk spacetime dimension $d$ such corrections can involve up to $d$ derivatives.  In particular, if the effective action happens to contain only an Einstein-Hilbert term at some cut-off energy $\Lambda$, then one-loop renormalization-group flow to a nearby scale $\Lambda -\Delta \Lambda$ will generally induce all terms with $n \le d$ derivatives with coefficients of order $G\Lambda^{d-n}$ (or $G\log \Lambda$ for $d=n$) relative to the Einstein-Hilbert term.  The contributions of such terms to computations at the scale $\Lambda$ are thus uniformly suppressed (up to logs) by the dimensionless parameter $G\Lambda^{d-2}$.  We thus consider general higher derivative terms below.

Below, we begin with a brief reminder (section \ref{sec:2derRev}) of certain features of fixed RT-area states and their flat saddle-point Renyi entropies derived in \cite{Dong:2018seb}.  Section \ref{sec:GAHD} then uses results from appendix \ref{app:LMVPRT} to rewrite this argument in an elegant form that (with help from appendix \ref{app:HDI}) allows ready generalization to the higher derivative case.

\subsection{Review of fixed RT-area states}
\label{sec:2derRev}

Suppose that we begin with a CFT state $|\psi\rangle$ defined by a Euclidean path integral, and that a Cauchy surface $\partial \Sigma$ for the CFT has been partitioned into regions $R$ and $\ol R$. As in \cite{Dong:2018seb}, for simplicity we take the state to be time-symmetric and the path integral to be real.  The AdS/CFT dictionary then defines a corresponding bulk path integral that computes the bulk wavefunction $\langle h | \psi \rangle$ where $|h\rangle$ is an eigenstate of the bulk induced metric on some bulk Cauchy surface $\Sigma$.  After gauge-fixing $\Sigma$ to run through the HRT-surface $\gamma_R$ and choosing coordinates on $\Sigma$ that fix the location of $\gamma_R$ on $\Sigma$, the bulk wavefunction $\langle h | \psi_{A_0}\rangle$ of the corresponding state $| \psi_{A_0} \rangle$ of fixed RT-area ${A_0}$ is defined by simply restricting $\langle h | \psi \rangle$ to metrics $h$ on $\Sigma$ that give $\gamma_R$ the desired area ${A_0}$.

\begin{figure}
\centering
\includegraphics[height=4cm]{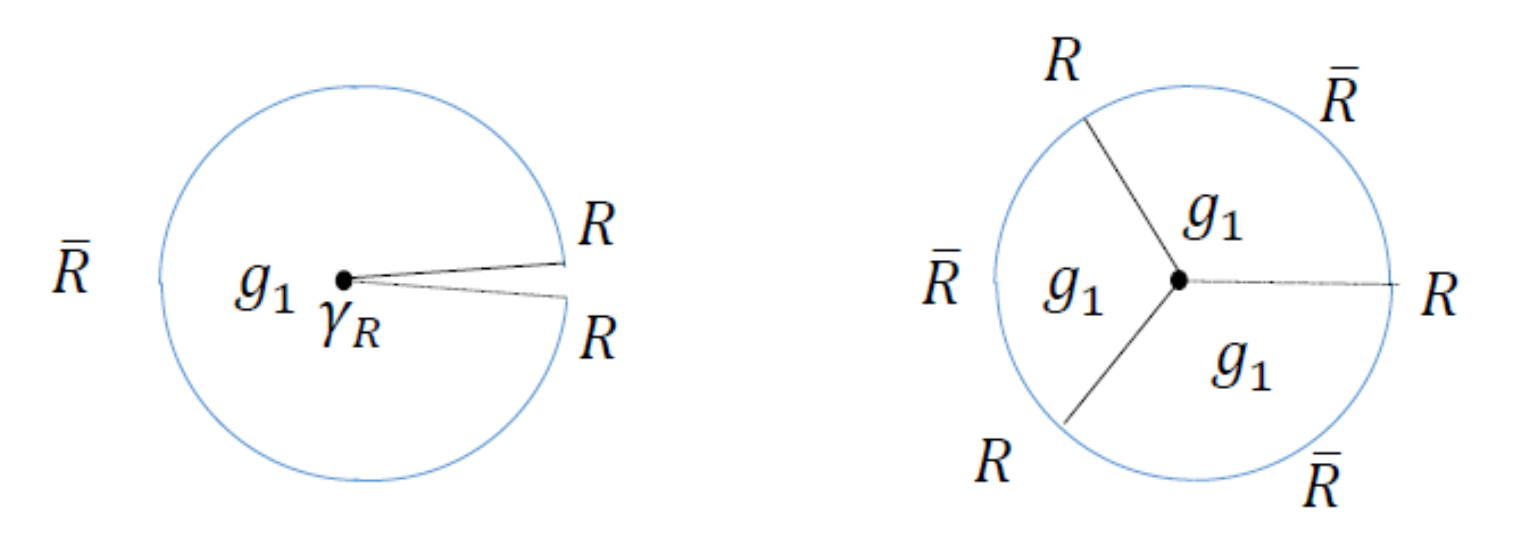}
\caption{After cutting open the $n=1$ bulk saddle $g_1$ (left), three copies may be glued together to construct the $n=3$ bulk saddle $g_{3}$ (right).  The black dot in the center is the HRT surface $\gamma_R$. }
\label{fig:Renyiglue}
\end{figure}

Note that the norm of $|\psi_{A_0}\rangle$ may be computed via
\begin{align}
\label{eq:fixAnorm}
\langle \psi_{A_0}  |\psi_{A_0}\rangle &= \int_h \ Dh\ |\langle h | \psi_{A_0}\rangle|^2\cr
&= \int_{\text{$h$ with area $A_0$ on $\gamma_R$}} \ Dh\ |\langle h | \psi\rangle|^2.
\end{align}
This is identical to the bulk path integral for $\langle\psi|\psi\rangle$ except that one treats the area of the HRT surface as fixed and not as a variable over which one integrates.  In the semiclassical limit, this means that allowed saddles $g_{A_0}$ for     \eqref{eq:fixAnorm} satisfy the same boundary conditions at AdS-infinity as saddles for $\langle\psi|\psi\rangle$, but that one of the bulk equations of motion fails to be enforced at $\gamma_R$.  The effect on the allowed solutions can be seen by introducing a term $\mu (A[\gamma_R] - A_0)$ into the action and treating $\mu$ as a Lagrange multiplier.  In Euclidean Einstein-Hilbert gravity, this allows the introduction of a conical defect on $\gamma_R$ whose magnitude is determined by the condition
$A[\gamma_R] = A_0$.  As a result, if $g_1$ is an allowed bulk saddle satisfying boundary conditions ${\cal B}_1$ as in figure \ref{fig:Renyiglue} (left), then for boundary conditions ${\cal B}_n$ given by simply sewing together $n$ copies of ${\cal B}_1$, we may construct an allowed bulk saddle $g_n$ by applying an analogous cut-and-paste procedure to
$g_1$ as in figure \ref{fig:Renyiglue} (right).

A direct calculation of saddle-point Renyi entropies or refined Renyi entropies \cite{Lewkowycz:2013nqa,Dong:2016fnf} then shows that they do not depend on $n$. In particular, defining the refined Renyi entropy $\wtd S_n(\rho_R)$ as the von Neumann entropy of the normalized density matrix $\rho_R^n/(\Tr \rho_R^n)$, this quantity is given by $A/4G$ for the RT surface associated with the saddle $g_n$.  Thus $\wtd S_n(\rho_R) = A_0/4G$ is constant.  But a short calculations also shows
\begin{equation}
\label{eq:refined}
\wtd S_n(\rho_R) = n^2 \partial_n \left( \frac{n-1}{n}S_n(\rho_R)\right).
\end{equation}
Integrating this relation for constant $\wtd S_n$ then gives $S_n = \wtd S_n = A_0/4G$, showing that the usual Renyi entropies are constant as well.

\subsection{Reformulation and higher derivative corrections}
\label{sec:GAHD}

We now wish to rewrite the above argument for constant Renyi entropies in a more elegant form that will generalize directly to actions with higher derivative corrections.  As stated above, our goal is to discuss classical variational principles for spacetimes with fixed geometric entropy determined by a region $R$ of their boundary.  And as noted above, at least for the two-derivative Einstein-Hilbert action the associated saddles involve conical singularities.  A complication, however, is that higher-derivative geometric entropy has not previously been studied carefully in spacetimes with non-zero conical defect angles.  We must thus not only construct an appropriate variational principle and find associated saddles, we must also determine what it means to fix geometric entropy in this context.

We propose that all of these questions be answered simultaneously by an appropriate analytic extension of the fixed-geometric-entropy action from cases where the results are clear.  We will show that such an analytic extension can be constructed by first considering variational principles for Euclidean spacetimes with codimension-2 conical defects with defect angles that are fixed as a boundary condition.  In the spirit of \cite{Carlip:1993sa}, we may then perform a Hamilton-Jacobi-like variation with respect to the defect angle boundary condition.  At vanishing defect angle, this latter variation is equivalent to the Lewkowycz-Maldacena computation of the entropy.  However, we may also perform this variation about backgrounds with non-zero conical defect angle and to thus define geometric entropy in those backgrounds. A Legendre transform then gives a variational principle appropriate to fixing this geometric entropy and simultaneously provides the analytic extension mentioned above.

Our starting point will be to observe that the Lewkowycz-Maldacena procedure for deriving the (two-derivative) Ryu-Takayanagi relation can be interpreted as just such a Hamilton-Jacobi-like variation of a fixed-conical-deficit action with respect to the conical deficit.  This is established in detail in appendix \ref{app:LMVPRT}.  In particular, we show there that the Einstein-Hilbert action provides a well-defined variational principle for an appropriate class of spacetimes with codimension-2 conical defects with fixed conical deficit angle $\delta$ so long as one ignores (a la Lewkowycz-Maldacena) the contribution to this action from the defect itself. This variational principle imposes Einstein's equations away from the defect {\it and also} imposes a natural analogue of the condition that the defect lie on an extremal surface.

It is useful to parametrize
the conical angle using a `replica number $m$' such that the opening angle at the defect is $2\pi m = 2\pi-\delta$; i.e., the defect-free case is $m=1$.  Even though we call it a replica number, $m$ can take any positive real value.   Note that this is a bulk replica number.  In contrast, in the Lewkowycz-Maldacena construction an integer boundary replica number $n$ leads to a quotient geometry in the bulk with opening angle $2\pi/n$ at the defect.  So their $n$ is related to our $m$ by $m=1/n$.  With this understanding, and assuming only minimal couplings of matter to gravity, the two-derivative geometric entropy $A_{\rm HRT}/4G$ is precisely the variation of the fixed-$m$ two-derivative action $\td I^{(2)}_m$ with respect to $m$ up to an overall sign:
\begin{equation}
\label{eq:I2sigma}
\frac{d \td I^{(2)}_m}{dm} = - \frac{A_{\rm HRT}}{4G}.
\end{equation}
Here the tilde in $\td I^{(2)}_m$ is meant to emphasize that it is the action for a fixed conical angle, to be distinguished with the fixed-geometric-entropy action that we will introduce later.  In particular, $\td I^{(2)}_m$ does not include any contribution from the conical defect itself.

Evaluating the result \eqref{eq:I2sigma} at $m =1$ (i.e., at $\delta =0$) gives a rewriting of the Lewkowycz-Maldacena derivation \cite{Lewkowycz:2013nqa} of the Ryu-Takayanagi entropy.  But in the above form the Lewkowycz-Maldacena argument now extends to saddles of the given action with general $m \neq 1$.  Passing to the Legendre transform simply adds a Lagrange multiplier that fixes $A_{\rm HRT}/4G$ to the desired value.  As discussed in \cite{Dong:2018seb}, this fixed-area action gives the leading semi-classical contribution to the partition function for states with the given value of $A_{\rm HRT}/4G$.

We now wish to repeat the steps in the above argument for actions with higher derivative corrections.  The key technical point is established in appendix \ref{app:HDI}, which shows that a recipe analogous to the no-singularity-contribution Einstein-Hilbert protocol above continues to define a good variational principle with action $\td I_m$ at fixed conical deficit $\delta = 2\pi(1-m)$ for general $m>0$ in the presence of perturbative higher derivative terms.

In this context,  ignoring contributions from the conical defect typically involves cancelling divergences with counter-terms (including some divergences that now arise from the Einstein-Hilbert term).  In other words, the singularity can be associated with contributions that are not just $\delta$-functions localized at the defect.  While it is thus not a priori clear how to divide such contributions into parts ``associated with the bulk'' and parts ``associated with the defect," we choose counter-terms that make the result analytic in the conical angle. One may thus also think of this procedure as analytic continuation from cases where counter-terms are not required, and in particular from the cases of integer $n=1/m$ where the spacetime admits a smooth $n$-fold cover so that the action may be defined as
\begin{equation}
\label{eq:quotient}
\td I_{m=\frac{1}{n}} : = \frac{1}{n}\td I_1(\text{$n$-fold cover}),
\end{equation}
with $\td I_1$ being the usual higher derivative action on smooth spacetimes.   At all $m$, repeating the Lewkowycz-Maldacena argument with this action then implies the geometric entropy to be
\be
\sigma = - \fr{d \td I_m}{dm},
\ee
in analogy with \er{eq:I2sigma}. Again, the variational principle imposes a condition that one may think of as placing the conical defect on a surface that extremizes the geometric entropy\footnote{\label{foot3}The recipe of appendix \ref{app:HDI} reproduces the standard definitions of both $\sigma$ and the action at $m=1$ ($\delta =0$).  As a result, at first order in $m-1$ our action $\td I_m$ may be written $\td I_m = \td I_1 - (m-1) \sigma$.  And again, the result gives a natural analytic extension of results that follow from \er{eq:quotient} when  $n=1/m$ is an integer and allows metric variations that may be interpreted as moving the surface on which $\sigma$ is evaluated relative to a smooth background geometry.  In this context, it is clear that varying $\td I_m$ about $m=1$ leads to a source of order $(m-1)$ on a surface extremizing $\sigma$ in the $m=1$ geometry.  Now, for more general $m$, the stationary points of $\td I_m$ have deficit angles $\delta = 2\pi(1-m)$ on the corresponding surface.  Taking ${\mathds Z}_n$ quotients then shows that $\sigma$ is extremized on shell whenever $1/m \in {\mathbb Z}$.  Finally, appendix \ref{app:HDI} derives a sense in which our construction analytically extends this condition to general real $m > 0$. See this appendix for details.}.

The variational principle appropriate to fixing the geometric entropy $\s$ may then be constructed as the Legendre transform
\begin{equation}
\label{eq:LT}
I_\sigma = \td I_m + (m-1) \sigma.
\end{equation}  Recall that Legendre transform gives the unique such action up to the addition of a function of $\sigma$. In \eqref{eq:LT} we have fixed this freedom by requiring consistency with the standard problem where $\sigma$ is unconstrained.  In that case the outcome $m=1$ may be thought of as an equation of motion.  Requiring the extremum of $I_\sigma$ with respect to $\sigma$ to match this by setting $m=1$ gives \eqref{eq:LT} up to the remaining freedom to add an overall constant.  As discussed in \cite{Dong:2018seb}, this is equivalent to noting that the condition $m=1$ selects the dominant value of $\sigma$ in the unconstrained problem.    This observation in turn means that $I_\sigma$ should agree with the standard higher derivative action $\td I_1$ when $m=1$ and removes the possibility of adding an overall constant to \eqref{eq:LT}.

We may now repeat the argument of \cite{Dong:2018seb} to show that the associated saddle-point Renyi entropies are flat, with $S_n^{\rm saddle}$ independent of $n$.  In particular, consider a CFT state $|\psi\rangle$ defined by a CFT path integral with sources.  Holographic duality allows the norm $\langle \psi |\psi\rangle$ to be computed using a bulk gravitational path integral with boundary conditions specified by the sources in the CFT path integral.  We further wish to consider the state $|\psi\rangle_\sigma$ defined by projecting
$|\psi\rangle$ onto a (perhaps approximate) eigenstate of the geometric entropy with eigenvalue $\sigma$.  In the saddle-point approximation the norm of $|\psi\rangle_\sigma$ is given by $e^{-I_\sigma[g_1]}$ where the above action $I_\sigma$ has been evaluated on a saddle point $g_1$ satisfying the above-mentioned boundary conditions at AdS infinity.

We wish to compute saddle-point Renyi entropies of $|\psi\rangle_\sigma$.  This means that we consider the CFT path integral defined by appropriately gluing together $n$ copies of the path integral for $|\psi\rangle$, and then study the corresponding bulk gravitational path integral with a constraint inserted to fix the geometric entropy to $\sigma$.  We define the associated saddle-point Renyi entropies $S^{\rm saddle}_n$ by approximating such path integrals by $e^{-I_\sigma[g_n]}$ evaluated on Euclidean solutions $g_n$ satisfying this constraint.

Such saddles $g_n$ are now easy to construct.  In the variational principle for fixed $\sigma$, the conical deficit $2\pi(1-m)$ is a dynamical variable chosen to obtain the specified geometric entropy.  It will thus vanish only for certain values of $\sigma$ for a given choice of $|\psi\rangle$. For our choice of $\sigma$ and the associated $g_1$, we let $\phi_{1} = 2\pi m_1$ denote the opening angle of the associated cone (so that the space is smooth only for $\phi_{1}=2\pi$ or $m_1=1$). The saddles $g_n$ are then found by cutting open $n$ copies of $g_1$ and sewing  them together as described in figure \ref{fig:Renyiglue} to give $m_n = nm_1$.

This $g_n$ clearly satisfies the desired boundary conditions at AdS infinity.  Furthermore, as shown in appendix \ref{app:HDI}, $\sigma[g_1]$ is fully determined by the properties of $g_1$ in the region near the defect. In particular, it may be computed by taking a limit as one approaches the conical defect from any fixed direction.  As a result, $\sigma[g_n] = \sigma[g_1]$ and the geometric entropy takes the desired value as well.  It follows that $g_n$ is a saddle for $I_\sigma$ satisfying all boundary conditions.  We shall assume such saddles to dominate in the path integral, though of course this issue deserves more study in the future.

Let us now discuss the value of $I_\sigma$ on $g_{n}$ following \cite{Dong:2018seb}.  We begin with the fundamental saddle $g_{1}$ and compute $I_\sigma[g_1] = \td I_{m_1}[g_1] + (m_1-1)\sigma$.  As stated above, the first contribution $\td I_{m_1} $ comes from ignoring contributions from the conical defect.  In particular, it can be obtained by cutting out a region of radius $\epsilon$  around the defect, including appropriate counter-terms at the new inner boundary, and taking the limit $\epsilon \rightarrow 0$.  Since $g_n$ consists of $n$ copies of $g_1$ away from the defect, this implies $\td I_{m_n}[g_n] = n \td I_{m_1}[g_1]$.  Since $m_n=nm_1$, the full action satisfies
\begin{equation}
\label{eq:Isiggn}
I_\sigma[g_n] = n \td I_{m_1}[g_1] + (nm_1 -1) \sigma.
\end{equation}
Taking into account proper normalization of $\rho_R$ then yields
\begin{equation}
\left( \log {\Tr} \rho_R^n  \right)^{\rm saddle}= \log \frac{e^{-I_\sigma[g_{n}]}}{e^{-nI_\sigma[g_{1}]}}  = - (I_\sigma[g_{n}] -nI_\sigma[g_{1}]) = -(n-1)\sigma,
\end{equation}
so that the saddle-point Renyi entropies defined by \eqref{eq:Renyi} yield $S_n^{\rm saddle} = \sigma$ and are indeed independent of $n$.  Alternatively, we could again have noted that the saddle-point refined Renyi entropies $\wtd S_n^{\rm saddle}$ are again fixed by the condition on $\sigma$ and then integrated \eqref{eq:refined} to find $S_n^{\rm saddle} = \sigma$ as well.

\section{Density matrix multiplication in states of fixed geometric entropy}
\label{sec:result}

We now have all the tools in hand to flesh out the argument sketched at the end of section \ref{sec:code} that multiplication by a code-subspace density matrix preserves code-subspaces with fixed geometric entropy.   We begin with a careful description of the appropriate code subspaces ${\cal H}_\phi$.  As discussed in footnote \ref{foot:Hphi}, the usual description of ${\cal H}_\phi$ as the space of states describing small quantum fluctuations about a given classical background is not consistent with the statement that it contains states of fixed geometric entropy $\sigma$, as any observable ${\cal O}$ that fails to commute with $\sigma$ will necessarily have significant fluctuations.  This is much like the statement in familiar non-relativistic quantum mechanics that position eigenstates allow large fluctuations in momenta.

Since we wish to work with such fixed-$\sigma$ states, it is useful to instead define ${\cal H}_\phi$ as the linear span of states constructed from fixed-$\sigma$ Euclidean path integrals using some given set of classical sources and arbitrary additional sources of order $1$ in counting powers of $G$; i.e., we consider small deviations from some given fixed-$\sigma$ state. 
States defined in this way allow what we may call large fluctuations in the conical angle at the defect, but fluctuations elsewhere are small.  Furthermore, the large fluctuations in conical angle need not obstruct semiclassical computations of ${}_\sigma \langle \psi | \psi \rangle_{\sigma}$, and indeed it is precisely such large fluctuations that allow the conical angle of a saddle to be tuned to satisfy the constraint on $\sigma$.  This is in direct parallel to the situation in non-relativistic quantum mechanics when using the semiclassical approximation to study the propagator $\langle x', t'|x,t\rangle$ between exact position eigenstates.

In making the above definition of ${\cal H}_\phi$, one should note that sources generally have non-trivial conformal dimensions so that the magnitude of any source depends on a choice of conformal frame.  Now, in considering Renyi entropies associated with the division of a CFT Cauchy surface into regions $R, \ol R$, the natural conformal frames to use are those in which the mutual boundary $\partial R = \partial \ol{R}$ of $R$, $\ol{R}$ has been pushed to infinity.  We have in mind such frames below.

In this context the arguments of \cite{Faulkner:2013ana} again imply an FLM formula on this ${\cal H}_\phi$, from whence Ref.~\cite{Almheiri:2014lwa,Dong:2016eik,Harlow:2016vwg} show states in ${\cal H}_\phi$ to be a code subspace\footnote{These arguments are typically made for finite-dimensional Hilbert spaces.  As stated above, we use a conformal frame where the CFT has non-compact Cauchy surfaces.  So even with a UV cutoff, the Hilbert space has infinite dimension.  We assume that the conclusion nevertheless continues to hold.  We presume this can be argued by first imposing and then removing a suitable IR regulator.} with the QEC structure described in section \ref{sec:code}.    It will remain useful to think of $\phi$ as a classical background (perhaps with a conical defect) and to take $\sigma$ to be the corresponding geometric entropy.

We now choose a state $|\psi\rangle_\sigma \in {\cal H}_\phi$ and consider the new state $|\psi_3\rangle_\sigma$ defined by multiplying  $|\psi \rangle_\sigma $ by the density matrix that it defines on $R$.  Specifically, we define
\begin{equation}
\label{eq:cube2}
|\psi_3\rangle_\sigma := \left(\rho_{R}  \otimes {\mathds 1}_{\ol{R}} \right) |\psi\rangle_\sigma = \left(e^{-K_{R}}  \otimes {\mathds 1}_{\ol{R}}\right) |\psi\rangle_\sigma,
\end{equation}
where $\rho_{R}$ is the density matrix defined on $R$ by tracing $|\psi\rangle_\sigma$ over $\ol R$ and $K_{R}$ is the associated modular Hamiltonian.  The state is labelled with a subscript $3$ because ${\Tr}_{\ol R} \ |\psi_3\>_\sigma {}_\sigma\<\psi_3| = \rho_R^3$ and thus
\begin{equation}
\label{eq:cubenorm}
{}_\sigma\<\psi_3|\psi_3\>_\sigma = {\Tr} \left(\rho_R^3\right).
\end{equation}

It will be useful to also consider the bulk operator $\left(\rho_{W}\otimes  {\mathds 1}_{\ol{W}} \right)$ defined by the density matrix for bulk quantum fields in the entanglement wedge $W$ of the CFT region $R$ induced by the global state $|\psi\>_\sigma$.  One may think of $\rho_{W}$ as defined by a bulk path integral for quantum fluctuations on the dominant classical saddle $g_1$ in the path integral computation of the norm ${}_\sigma\<\psi|\psi\>_\sigma$, after cutting this path integral open along the slice defined by the ${\mathds Z}_2$ symmetry that exchanges (and complex conjugates) corresponding sources associated with the bra- and ket-vectors.  As a result, acting with $\left( \rho_{W} \otimes {\mathds 1}_{\ol{W}} \right)$ extends a bulk path integral for quantum fluctuations by splicing in a copy of $g_1$ in much the same manner that inserting two copies of $\left(\rho_{R} \otimes  {\mathds 1}_{\ol{R}} \right)$  into the path integral for
${}_\sigma\<\psi|\psi\>_\sigma$ extends it to the 3-replica path integral for \eqref{eq:cubenorm}.  Indeed, at this order in the bulk semiclassical approximation, the only difference between insertions of these two operators is that the latter also changes the classical contribution $e^{-I_\sigma}$ while the former does not.

We can now use this observation to show that $|\psi_3\rangle_\sigma$ lies in the same code subspace ${\cal H}_\phi$.  We will proceed by proving that the results of acting on $|\psi\rangle_\sigma$ with either
$\left(\rho_{W}\otimes {\mathds 1}_{\ol{W}}  \right)$ or $\left(\rho_{R}  \otimes {\mathds 1}_{\ol{R}} \right)$  yield identical states up to an overall normalization and  corrections that can be neglected at our one-loop level.
In particular, let us define
\be
|\psi_3{}'\rangle_\sigma := \left( \rho_{W} \otimes {\mathds 1}_{\ol{W}}  \right) |\psi\rangle_\sigma = \left(e^{-K_{W}}  \otimes {\mathds 1}_{\ol{W}} \right) |\psi\rangle_\sigma,
\ee
where $\r_{W}$ and $K_{W}$ are the bulk density matrix and bulk modular Hamiltonian defined on the entanglement wedge of $R$ by $|\psi \rangle_\sigma$.  We also recall the construction of $n$-replica saddles $g_n$ defined as in figure \ref{fig:Renyiglue} by cutting and sewing copies of $g_1$.   Assuming as in section \ref{sec:HD} that $g_n$ is the dominant saddle in the computation of ${\Tr } \rho_R^n$,  the above observations imply
\be
\label{eq:33}
{}_\sigma\<\psi_3|\psi_3\>_\sigma = e^{-I_\sigma[g_3]} Z_{\rm bulk \ flucts}[g_3],
\ee
\be
\label{eq:3p3p}
{}_\sigma\<\psi_3{}'|\psi_3{}'\>_\sigma  = e^{-I_\sigma[g_1]} Z_{\rm bulk \ flucts}[g_3],
\ee
and
\be
\label{eq:33p}
{}_\sigma\<\psi_3{}|\psi_3{}'\>_\sigma = e^{-I_\sigma[g_2]} Z_{\rm bulk \ flucts}[g_3],
\ee
at all orders in $G$.

The important observation above is then that since all three inner products involve two insertions of $\rho_R$, two insertions of $\rho_{W}$, or one of each, in each case $Z_{\rm bulk \ flucts}$ is evaluated on the same 3-replica saddle $g_3$.  As a result, these contributions cancel when computing
\be\la{overlap}
\fr{{}_\sigma \< \psi_3|\psi_3{}'\>_\sigma^2}{{}_\sigma\<\psi_3|\psi_3\>_{\sigma} {}_\sigma\<\psi_3{}'|\psi_3{}'\>_{\sigma}} = e^{-2I_\sigma[g_2] + I_\sigma[g_1] + I_\sigma[g_3]} = 1,
\ee
where the last equality follows from the linearity in $n$ of $I_\sigma[g_n] = n \td I_{m_1} [g_1] - (nm_1-1)\sigma$ in \eqref{eq:Isiggn}. Equation \eqref{overlap} should be understood to hold to all orders in $G$, though there are non-perturbative corrections due to sub-leading saddles in \eqref{eq:33}--\eqref{eq:33p}. 

As $\rho_{W}$ is a bulk operator that acts within the code subspace ${\cal H}_\p$, the state $|\psi_3'\>$ must lie in ${\cal H}_\p$.  We thus see that $\left(\rho_{R} \otimes {\mathds 1}_{\ol{R}}   \right) |\psi\rangle_\sigma$ lies in ${\cal H}_\phi$ up to small corrections as claimed.  Furthermore, as described at the end of section \ref{sec:code}, this in turn requires $ \left(\chi_{R_\alpha^2}\right)^3 \propto \chi_{R_\alpha^2}$  (equation \eqref{eq:chicube}) for each superselection sector $\alpha$ that appears in ${\cal H}_\phi$. And since density matrices have real non-negative eigenvalues, the eigenvalues of $\chi_{R_\alpha^2}$ can be only $0$ and $1$ up to an overall normalization.  We thus conclude that $\chi_{R_\alpha^2}$ is a projector onto a subspace of dimension dictated by its entropy, and that multiplication by
$\left(\rho_{R}  \otimes {\mathds 1}_{\ol{R}} \right)$ leaves ${\cal H}_\phi$ invariant\footnote{This follows from the decomposition \eqref{basisrep} and the identity \eqref{eq:chicube}, though one can also generalize the above argument directly to the case where $\rho_R$ is the density matrix of a distinct state $|\psi'\rangle_\sigma  \neq |\psi\rangle_\sigma\in {\cal H}_\phi$.}.

\section{Discussion}
\label{sec:disc}

As reviewed in section \ref{sec:code}, at $O(1)$ in the bulk Newton constant $G$, holographic quantum codes allow complementary recovery and thus are characterized up to unitaries by the spectrum of a class of density matrices called $\chi_{R_\alpha^2}$, where $\alpha$ labels superselection sectors with respect to the bulk algebra recovered by the code.  Our arguments above used properties of bulk gravitational path integrals to show, again 
up to higher order $O(G)$ corrections, that
 each $\chi_{R_\alpha^2}$ is proportional to a projection operator of rank determined by the geometric entropy $\sigma = A/4G + \dots$ (with $\dots$ denoting higher derivative terms) associated to the given superselection sector $\alpha$. Here we measure the magnitude of any corrections  by their impact on the Renyi entropies $S_n(\rho_R) \equiv - \frac{1}{n-1} \log {\Tr} (\rho_R^n),$ taking $n$ fixed in the limit of small $G$.
 
Because the non-zero eigenvalues $\lambda_k$ of $\chi_{R_\alpha^2}$ are independent of $k$ up to the stated corrections, we refer to this result as one-loop flatness of the entanglement spectrum for holographic quantum codes.  Our arguments apply to gravitational systems where the effective action is Einstein-Hilbert plus matter with arbitrary perturbative higher derivative corrections, such as those controlled by small $\alpha'$ or $G$.   In parallel with past assumptions \cite{Lewkowycz:2013nqa,Dong:2018seb} that any breaking of replica symmetry is subdominant, our arguments assume that saddles of the form shown in figure \ref{fig:Renyiglue} dominate the relevant path integrals.  It would clearly be of use to explore this assumption more completely in future work.
 
An important technical step (see appendices \ref{app:LMVPRT} and \ref{app:HDI}) was to construct good variational principles that even in the presence of arbitrary higher derivative corrections allow spacetimes with conical defects, and to show (see section \ref{sec:HD}) that the geometric entropy is given by a Hamilton-Jacobi-like variation of the on-shell action $\tilde I_m$ with respect to the defect angle.  This in particular identifies the Lewkowycz-Maldacena procedure \cite{Lewkowycz:2013nqa} as the natural analogue of the Carlip-Teitelboim approach to black hole entropy \cite{Carlip:1993sa} generalized to cases that lack the $U(1)$ symmetry of \cite{Carlip:1993sa}.  It also further develops the machinery of higher-derivative corrections for use in other applications, and in particular provides an appropriate analogue at finite conical angle of the extremality condition for the geometric entropy $\sigma$.

As in \cite{Dong:2018seb}, the fact that our defects are spacelike means that the corresponding path integrals prepare states of the original defect-free theory, and in particular that the defect makes no contribution to the Hamiltonian or momentum constraints on any Cauchy surface $\Sigma$ passing through the defect.  This follows from the fact that the defect in no way constraints the lapse and shift on $\Sigma$, and from the fact that (since it is a geometric invariant) the counter-term defined in appendix \ref{app:HDI} can be constructed from canonical data on $\Sigma$ without involving either lapse or shift.  Integrating over lapse and shift thus imposes the defect-free constraints as in \cite{Halliwell:1990qr}, though here including appropriate higher derivative corrections.

Our one-loop flatness for $\chi_{R_\alpha^2}$ provides a useful extension of the results of \cite{Akers:2018fow,Dong:2018seb}, which showed {\it any} semi-classical bulk state to have flat entanglement spectrum at $O(1/G)$.  It is not possible that such a strong result holds at $O(1)$ since one can use the dynamical IR quantum fluctuations to engineer by hand a state (on the whole system) with non-flat entanglement spectrum at $O(1)$, but we identify the essence of the result as relating to the structure of the quantum code rather than to individual encoded states.

The above universal form of holographic codes matches well with that found in simple tensor network models \cite{Pastawski:2015qua,Hayden:2016cfa}, with the caveat that such models should be interpreted as describing states of fixed geometric entropy.  As described in \cite{Donnelly:2016qqt},
such models can be extended to so-called edge-mode tensor networks which describe more general states.  

As noted in \cite{Dong:2018seb}, and as we now briefly review, one-loop flatness of $\chi_{R_\alpha^2}$ also immediately implies a stronger version of the JLMS relation \cite{Jafferis:2015del} between boundary and bulk modular Hamiltonians than has previously been derived.  A CFT density matrix $\rho_R$ in a subregion $R$ defines a so-called boundary modular Hamiltonian $K_R = - \log \rho_R$, and on the Hilbert space ${\cal H}_\phi$ associated with bulk quantum fluctuations around a given classical background the bulk density matrix $\rho_W$ in the corresponding bulk entanglement wedge defines an analogous bulk modular Hamiltonian $K_W = -\log \rho_W$. JLMS showed $K_R$ to be related to $K_W$ and the area operator $A$ on the RT/HRT surface in a manner that is often written

\begin{equation}
\label{eq:UJLMS}
K_R = \frac{A}{4G} + K_W + O(G).
\end{equation}
However, it is important to recall that the argument \cite{Jafferis:2015del} for \eqref{eq:UJLMS} involves taking expectation values in code subspace states ${\cal H}_\phi$ associated with small fluctuations about a given classical solution\footnote{It is interesting to consider varying the FLM relation under $\rho_R \rightarrow \rho_R + \epsilon \delta \rho_R$ where $\rho_R, \delta \rho_R$ are associated with distinct ${\cal H}_\phi$, ${\cal H}_{\phi'}$.  The FLM relation generally describes changes in the associated entropy $S(\rho_R)$ to order $G^0$.  But in such cases this accuracy may not suffice to study very small values of $\epsilon$, where the changes in $S(\rho_R)$ can be exponentially small. So the full argument of \cite{Jafferis:2015del,Dong:2016eik} holds only within some fixed ${\cal H}_\phi$.}.   As a result, as stressed in \cite{Dong:2016eik} (but using the notation of \cite{Dong:2018seb}) the conclusion is best written
\begin{equation}
\label{eq:PJLMS}
P_C K_R P_C  = \(\frac{A}{4} + K_W\) P_C + O(G),
\end{equation}
where $P_C$ denotes the projection onto the appropriate ${\cal H}_{\phi}$. 

Although $K_R$ is naturally defined as an operator on ${\cal H}_R$,
following \cite{Dong:2016eik,Dong:2018seb} we use $K_R$ here and below to denote the operator $K_R \otimes {\mathds 1}_{\ol R}$ involving the identity ${\mathds 1}_{\ol R}$ on $\ol R$ and thus to define an operator on the full CFT Hilbert space.  We similarly use $K_W$ to denote $K_W \otimes {\mathds 1}_{\ol W}$.  On the right-hand side of \eqref{eq:PJLMS}, since both $A$ and $K_W$ are semi-classical bulk operators in the bulk effective field theory with a cutoff, they preserve any ${\cal H}_{\rm \phi}$ and so commute with $P_C$.  Thus it is sufficient to have a single $P_C$ on the right. We refer to \eqref{eq:UJLMS} as the unprojected JLMS relation in contrast to the projected relation \eqref{eq:PJLMS}.

A corollary of our one-loop flatness argument is that the stronger version of the JLMS relation \eqref{eq:UJLMS} does in fact hold when both sides are viewed as operators on the given ${\cal H}_{\phi}$. As noted in \cite{Dong:2018seb}, since \eqref{eq:PJLMS} is already known to hold this is equivalent to the statement that $P_C$ commutes with $K_R = - \log \rho_R$ , and thus also to the statement that multiplication by $\rho_R\otimes {\mathds 1}_{\ol{R}}$ preserves ${\cal H}_{\phi}$.  But that is precisely what was shown in section \ref{sec:result}.  Our result is also equivalent to the requirement that the boundary modular flow induced by $K_R$ preserves ${\cal H}_{\phi}$ (i.e., $e^{-iK_Rs}$ acts within ${\cal H}_{\phi}$), or equivalently
\begin{equation}
\label{eq:modflow}
    e^{-iK_Rs} P_C e^{iK_Rs} = P_C.
\end{equation}
As a result, establishing \eqref{eq:UJLMS} may allow greater use of modular flow in AdS/CFT\footnote{Nonetheless, see \cite{Faulkner:2017vdd,Chen:2018rgz,Faulkner:2018faa} for important applications thus far.  In particular, as noted in \cite{Dong:2018seb}, while \eqref{eq:PJLMS} does not generally imply a useful relation between bulk and boundary modular flows of arbitrary operators, it does suffice for modular flows of operators in $R$ that reconstruct bulk operators.  In particular, it suffices for the algorithm described in \cite{Faulkner:2017vdd}.}.

The fact that our code subspace ${\cal H}_{\rm \phi}$ is invariant under multiplication by $\left(\rho_{R} \otimes  {\mathds 1}_{\ol{R}}\right)$ and thus by $\left(\rho_{R}^{is} \otimes {\mathds 1}_{\ol{R}} \right)$ also immediately implies that the set of density matrices on $R$ defined by ${\cal H}_{\rm \phi}$ is invariant under the modular flow induced by any such state.  Again, this extends an $O(1/G)$ result from \cite{Dong:2018seb}.   A related invariance of ${\cal H}_{\rm code} = \oplus_\phi {\cal H}_\phi$ was conjectured on physical grounds in \cite{Faulkner:2017vdd}.  A relevant comment here is that we defined ${\cal H}_\phi$ so as to allow rather large bulk IR effects at $O(1)$ at a level analogous to allowing finite temperature states in flat-space quantum field theory.  Effects of this size are usually thought of as taking one outside of the Hilbert space that contains the Minkowski vacuum. In our context, in a conformal frame where these bulk IR effects correspond to the UV in the dual CFT, they may similarly take one outside the natural Hilbert space of states with good CFT duals.  However, this issue can be cured by imposing appropriate UV/IR cutoffs.  It would also be interesting to return to this issue using the technology of \cite{Kang:2018xqy}.

To discuss further interpretations and implications of our results, recall from section \ref{sec:code} that holographic quantum codes should in fact be viewed as families of codes labelled by an energy scale $\Lambda$.  At each $\Lambda$, properties of the code are determined by the bulk effective action at this scale, and the RG flow of the action must induce an associated RG flow of the quantum code.  Furthermore, because properties of the quantum fluctuations at scales below $\Lambda$ define the state to be encoded,  it is natural to expect that the manner in which such states are encoded is determined by other aspects of the bulk theory, and in particular by saddle-point computations involving the bulk effective action at the scale $\Lambda$.  This would mean that such calculations would determine the properties of $\chi_{R_\alpha^2}$.  Our work supports this conjecture, as we found in section \ref{sec:HD} that Renyi entropies computed using only saddle-point contributions to the gravitational path integrals would indeed match the above results for $\chi_{R_\alpha^2}$.

Since the form of the effective action and thus the quantum code generally vary with $\Lambda$, the
universal form of our result requires interesting cancellations to occur between various aspects of the associated RG flow.  In particular, shifting $\Lambda$ generates changes in the effective action by integrating out additional degrees of freedom under the assumption that they remain in their local vacuum state.  But vacuum states are known to have thermal spectra and, especially in a context with a large-$N$ matter sector where there are many bulk matter fields but only a single graviton, constraining the geometric entropy will have little effect on this thermal result.  Thermal spectra are not flat, but have a Boltzmann distribution of eigenvalues.  So a coarse graining that simply reorganizes vacuum dynamical quantum fluctuations from $\mathcal{H}_{R_\alpha^1}, \mathcal{H}_{\ol{R}_\alpha^1}$ by absorbing them into the states $|\chi_\alpha\ran \in \mathcal{H}_{R_\alpha^2} \otimes \mathcal{H}_{\ol{R}_\alpha^2}$ would violate flatness at one-loop order.  Such effects must thus cancel against others, perhaps associated with the fact that superselection operators like the geometric entropy $\sigma$ evolve with $\Lambda$, so that changes in the decomposition \eqref{eq:hilbertdecomp} itself must also be taken into account.  Indeed, the geometric entropy evolves in at least two ways as its explicit form depends on couplings in the effective action at the scale $\Lambda$ and also because $\sigma$ should be understood as being smeared over length scales of order $1/\Lambda$ in directions transverse to the RT surface.  It would be interesting to study such effects using either bulk gravitational path integrals or tensor network models.

As a final comment, we mention that we described our main results as being valid at one loop because we relied on the framework of QEC with complementary recovery.  Beyond one-loop order, the complementary recovery aspect is expected to break down, though some notion of QEC may remain; see e.g.\ the discussion in the final paragraph of \cite{Harlow:2016vwg}.  However, our intermediate results concerning variational principles and the bulk computation \eqref{overlap} in fact remain valid at arbitrary orders in perturbation theory.  As a result, they will continue to constrain whatever structures remain at higher orders.  Improving the understanding of QEC and higher order corrections is thus an important goal for future work.

\paragraph{Acknowledgments}
It is a pleasure to thank Adam Levine and Daniel Harlow for useful conversations.  This material is based upon work supported by the Air Force Office of Scientific Research under award number FA9550-19-1-0360.  DM were supported in part by the Simons Foundation.  XD and DM were also supported in part by funds from the University of California.  This work was developed in part at the KITP which is supported in part by the National Science Foundation under Grant No.\ PHY-1748958.

\appendix
\section{The Lewkowycz-Maldacena argument as a Hamilton-Jacobi variation}
\label{app:LMVPRT}

In this appendix, we work with Euclidean Einstein gravity and show that for metrics with a fixed conical defect angle, the Einstein-Hilbert action without including any contribution from the conical defect leads to a well-defined variational principle.  We find that the Hamilton-Jacobi variation of such an action with respect to the conical defect angle is determined by the area of the conical defect.  The Lewkowycz-Maldacena argument for computing the gravitational entropy can be interpreted as the special case of performing this Hamilton-Jacobi variation about backgrounds with vanishing conical deficit.  We also explicitly construct solutions to Einstein's equations with a general conical defect angle in a  systematic expansion valid near the defect and show that the trace of the analogue of the extrinsic curvature tensor vanishes on the defect.  Under appropriate asymptotic boundary conditions, the solution is generically unique up to residual gauge transformations.

Let us start by defining a suitable space of (generally off-shell) metric configurations that contain a conical defect on a codimension-2 surface with opening angle $2\pi m$, so that smooth spacetimes have $m=1$.  Here $m$ is any positive real number and not necessarily an integer.  We will work in a convenient set of quasi-cylindrical coordinates~\cite{Unruh:1989hy} defined by constructing normal geodesics from the conical defect, where the metric can be taken to be of the form
\be\la{met}
ds^2 = dr^2 + \[m^2+\hat{o}(r)\] r^2 d\p^2 + O(r^0) dy^i dy^j + O(r^2) d\p dy^i
\ee
near $r=0$, the location of the conical defect.  Here $\p$ is an angular coordinate taking values in $[0,2\pi)$, the $y^i$ denote an arbitrary set of coordinates on the conical defect, and we have introduced the notation $\hat{o}(r)$ to denote terms that vanish as $r \rightarrow 0$ at least as fast as some power law $r^\eta$ with $\eta > 1$.  The $\hat{o}(r)$, $O(r^0)$, and $O(r^2)$ terms generally depend on all coordinates $(r,\p,y^i)$ -- although due to the required periodicity under $\p\sim\p+2\pi$ they can be expanded as a Fourier series using integer powers of $e^{i\p}$.

Below, we first show in section \ref{subsec:varP} that the above action gives a good variational principle for the class of metrics \eqref{met} with fixed $m$. We then argue in section \ref{subsec:gen} that Einstein's equations indeed admit solutions compatible with \eqref{met}, and in fact do so with a particular form for the expansion around $r=0$ (so that, if desired, our variational principle could then be further restricted to metrics of this asymptotic form).  Furthermore, assuming this expansion, the equations of motion impose a condition that generalizes the extremal surface condition satisfied by RT surfaces at $m=1$.  Finally, we give a counting argument in section \ref{subsec:unique} to show that the freedom in such solutions is precisely what one expects to need to match general boundary conditions at large $r$.  In other words, we show that there are enough solutions of the form \eqref{met} to describe the expected physics, and we also show that matching solutions of the more specific asymptotic form described in section \ref{subsec:gen} to given large-$r$ boundary conditions will generally leave no continuous free parameters.  Instead, such solutions form a discrete set as one expects of a good non-linear elliptic boundary value problem.

\sse
{Variational principle}
\label{subsec:varP}

We define an action $\td I[g]$ for these metric configurations as simply the Einstein-Hilbert action (with a cosmological constant\footnote{This should not be confused with our UV cutoff called $\Lambda$ in the main text.} $\L$) but without including any contribution from the conical defect:
\be\la{ig}
\td I[g] = -\fr{1}{16\pi G} \lim_{\e\to0^+} \int_{r\ge\e} d^{d+1}x \sqrt{g} (R-2\L).
\ee
Here the total dimension is $d+1$ and $x=(r,\p,y^i)$ denotes the collection of all coordinates.  If the spacetime has boundaries (other than $r=0$), such as an asymptotically AdS boundary at $r=\infty$, the action \er{ig} should be supplemented by the standard boundary terms there although we do not write them explicitly.

We use a tilde on the left-hand-side of \eqref{ig} to emphasize that we simply integrate the Lagrangian down to $r=0$ and do not include any delta-function contribution or Gibbons-Hawking-York boundary term at the defect. In particular, \eqref{ig} coincides with the prescription for computing actions in conical defect spacetimes
used by Lewkowycz and Maldacena in \cite{Lewkowycz:2013nqa}.  In their case, for $\frac{1}{m} \in {\mathbb Z}$ the prescription followed from the fact that they actually wished to study the action of the smooth $\fr{1}{m}$-fold cover, and for $\frac{1}{m} \notin {\mathbb Z}$ it then followed by analytic continuation.  In contrast, we wish to directly study metrics with conical singularities for which the opening angle $2\pi m$ is fixed as a boundary condition.  However, the connections with geometric entropy described in \cite{Lewkowycz:2013nqa} inspire us to conjecture that \eqref{ig} provides a good variational principle for our problem.  This conjecture will be verified below.

The first step is to note that the $\e\to0^+$ limit in \er{ig} does in fact converge for the metric configurations \eqref{met}.  To see this, note that in the $(z,\zb,y^i)$ coordinates defined by $z=r e^{im\p}$, the metric \er{met} can be written
\bg\la{mz}
ds^2 = dzd\zb + T \fr{(\zb dz-z d\zb)^2}{z\zb} + h_{ij} dy^i dy^j + 2iU_j dy^j (\zb dz-z d\zb),\\
\la{uvwc}
T=\hat{o}(r),\qu
h_{ij}=O(r^0),\qu
U_j=O(r^0)
\eg
where $T$, $h_{ij}$, and $U_j$ are functions of all coordinates $(z,\zb,y^i)$.  In these coordinates, we have
\be
g_{\m\n} = g_{\m\n}\Big|_{r=0} + \hat{o}(r),\qu
\T\G{\r}{\m\n} = \fr{\hat{o}(r)}{r},\qu
\T R{\m}{\n\r\s} = \fr{\hat{o}(r)}{r^2}.
\ee
In particular, the Ricci scalar $R=\hat{o}(r)/r^2$ is locally integrable near $r=0$.  Thus the action $\td I[g]$ is finite, assuming that any potential divergences near asymptotic boundaries have been dealt with by the standard counterterms.

We now show that the action $\td I[g]$ leads to a well-defined variational principle under the boundary condition that fixes $m$ (or equivalently the conical angle).  Under a general, infinitesimal variation $\d g_{\m\n}$ of the metric, the action changes by
\bm\la{dig}
\d \td I[g] = \fr{1}{16\pi G} \lim_{\e\to0^+} \[\int_{r\ge \e} d^{d+1}x \sqrt{g} (G^{\m\n}+\L g^{\m\n}) \d g_{\m\n} \rt. \\
+ \lt.\lt. \int_\pa d^dX \sqrt{\g} n^{\m} (\nabla^\nu \d g_{\m\n} - \nabla_\mu \d \T g\n\n) \rt|_{r=\e} \],
\em
where the first integral is a bulk term that vanishes if the equation of motion is satisfied (setting the Einstein tensor $G^{\m\n}$ to $-\L g^{\m\n}$ in this case), and the second integral is a boundary term at $r=\e$.  Here $X=(\p,y^i)$ and $\g$ denote the coordinates and determinant of the induced metric on this codimension-$1$ boundary, while $n^\m$ is the unit normal vector in the $r$ direction.

In order to have a well-defined variational principle, the boundary term in \er{dig} must vanish for metric variations that preserve $m$.  To see that this is the case, note that in the $(z,\zb,y^i)$ coordinates we have
\be
\d g_{\m\n} = \d g_{\m\n}\Big|_{r=0} + \hat{o}(r),\qu
\na_\r \d g_{\m\n} = \fr{\hat{o}(r)}{r},
\ee
as long as $\d g_{\m\n}$ preserves $m$.  This, together with $\sqrt{\g}\sim r$ and $n^\m \sim r^0$, shows that the boundary term in \er{dig} vanishes as $o(\e)$ as $\e\to0$.

In addition, we note that varying the on-shell action with respect to $m$ gives the area of the conical defect, for any value of $m$.  To see this, note that in \er{dig} the bulk term vanishes if the equation of motion is satisfied, but the boundary term may be nonzero for a metric variation that changes $m$.  Working for example in the $(r,\p,y^i)$ coordinates, we find
\be
\lt.\lim_{\e\to0^+} \sqrt{\g} n^{\m} (\nabla^\nu \d g_{\m\n} - \nabla_\mu \d \T g\n\n) \rt|_{r=\e} = \lim_{r\to0^+} \sqrt{\g} \T\G{r}{\p\p} \d g^{\p\p} = -2\d m \sqrt{\bar h}
\ee
where $\bar h$ is the determinant of the induced metric $\bar h_{ij} \eq h_{ij}\big|_{r=0}$ on the conical defect.  Therefore,
\be
\fr{d \td I_m}{d m} = -\fr{1}{4G} \int d^{d-1}y \sqrt{\bar h} = -\fr{A}{4G},
\ee
where $\td I_m$ denotes the on-shell action with the boundary condition set by $m$.

\sse{General solutions}
\label{subsec:gen}

We wish to show that the metric ansatz \er{mz} allows general solutions to Einstein's equations. In particular, we now show that one can solve the equations of motion with functions $T$, $U_i$, and $h_{ij}$ having expansions near $r=0$ of the form
\ba\la{expt}
T &= \sum_{\substack{p,q,s=0\\pq>0 \text{ or } s>0}}^{\infty} T_{pqs} z^{\fr{p}{m}} \zb^{\fr{q}{m}} (z\zb)^s,\\
U_i &= \sum_{\substack{p,q,s=0}}^{\infty} U_{i,pqs} z^{\fr{p}{m}} \zb^{\fr{q}{m}} (z\zb)^s,\\
\la{expg}
h_{ij} &= \sum_{\substack{p,q,s=0}}^{\infty} h_{ij,pqs} z^{\fr{p}{m}} \zb^{\fr{q}{m}} (z\zb)^s.
\ea
Here the coefficients $T_{pqs}, U_{i,pqs}, h_{ij,pqs}$ can be arbitrary functions of the $y^i$.
Such a solution is manifestly periodic under $\p\sim\p+2\pi$ (or equivalently $z\sim z e^{2\pi im}$) as required.
In the present subsection we show only that the above expansions are consistent with the equations of motion, and that those equations impose a generalization of the extremal surface condition satisfied by RT surfaces at $m=1$.  We will return to the issue of whether they admit sufficiently general such solutions in section \ref{subsec:unique}.

Note that if $m$ happens to be a rational number, the expansions \er{expt}--\er{expg} involve redundant terms.  We will first study the generic case where $m$ is irrational, and then obtain results for rational $m$ by taking limits of the generic case.  We will find such limits to be well-behaved in Einstein gravity.

Let us start with the generic case where $m$ is an irrational number.  To see that \er{expt}--\er{expg} can consistently solve Einstein's equations, we first introduce some terminology.  We say that a function $f$ is of type $[\a]$ if it satisfies all three conditions below: 
\begin{enumerate}[label=\arabic*)]
\item At $r=0$ it has the expansion
\be\la{fexp}
f = \(\fr{z}{\zb}\)^{\ell/2} \sum_{\substack{p,q,s=0}}^{\infty} f_{pqs} z^{\fr{p}{m}} \zb^{\fr{q}{m}} (z\zb)^{s-\a},
\ee
with some integer $\ell$ (which we will call the angular momentum) and some $\a$ such that $\a+\ell/2$ is an integer. Note that both $\ell$ and $\a$ can have either sign, and that $\a$ is either integer or half-integer.

\item Nonzero terms in the expansion \er{fexp} do not have negative integer powers of $z$ or $\zb$.  In particular, $f_{pqs}$ vanishes if $p=0$ and $s-\a+\ell/2<0$ or if $q=0$ and $s-\a-\ell/2<0$.   This condition has the nice property that it is preserved by derivatives, additions, and multiplications.
\item Each coefficient $f_{pqs}$ is determined by the coefficients $T_{p'q's'}$, $U_{i,p'q's'}$, and $h_{ij,p'q's'}$ at lower orders, by which we mean
\be
\label{<def}
p'\leq p,\qu
q'\leq q,\qu
s'\leq s,\qu
(p',q',s') \neq (p,q,s).
\ee
We will use
\be
(p',q',s')<(p,q,s)
\ee
to denote the full set of conditions \eqref{<def}.
\end{enumerate}
Using an overline to indicates a form of closure, not complex conjugation, we will also say that
a function is of type $[\ol\a]$  if it fulfills conditions 1) and 2) above but, instead of 3), it satisfies the following variant:
\begin{enumerate}
\item[$\ol{\rm 3}$)] Each coefficient $f_{pqs}$ only depends on $T_{p'q's'}$, $U_{i,p'q's'}$, and $h_{ij,p'q's'}$ of lower or equal orders, by which we mean
    \be
    \label{pqsle}
(p',q',s') \leq (p,q,s) \qu
\iff\qu
p'\leq p,\qu
q'\leq q,\qu
s'\leq s.
\ee
\end{enumerate}

Let us now find a few useful properties of these two types of expansions.  We will think of $[\a]$ as a set and write
\be
f \in [\a]
\ee
if $f$ is of type $[\a]$, and similarly for $[\ol\a]$.  A function of type $[\a]$ is also a function of type $[\ol\a]$ which is in turn of type $[\a+1]$:
\be
[\a] \subset [\ol\a] \subset [\a+1].
\ee
Any two such functions with the same angular momentum $\ell$ can be added to yield a function with a similar expansion.  Using $[\a,\ell]$, $[\ol{\a,\ell}]$ to denote functions of type $[\a]$, $[\ol{\a}]$ with angular momentum $\ell$, we may thus write
\be
[\a,\ell]+[\b,\ell] \subset [\max(\a,\b),\ell],\qu
[\ol{\a, \ell}]+[\ol{\b, \ell}] \subset [\ol{\max(\a,\b),\ell}].
\ee
We also have
\ba
&z,\, \zb \in [-1/2], \\
&\pa_a [\a] \subset [\a+1/2],\qu
\pa_a [\ol\a] \subset [\ol{\a+1/2}],\\
&\pa_i [\a] \subset [\a],\qu
\pa_i [\ol\a] \subset [\ol\a],
\ea
where indices such as $a$ denote either $z$ or $\zb$.  For products we have the general rules
\be
[\a] [\b] \subset [\a+\b],\qu
[\a] [\ol\b] \subset [\ol\a] [\ol\b] \subset [\ol{\a+\b}],
\ee
as well as three special rules associated with the subset $[\a]^+ \subset [\a]$ defined to contain precisely those functions $f \in [\a]$ with $f_{000}=0$ (where $f_{000}$ is defined using the expansion \eqref{fexp} with the given $\a$ value\footnote{Since $[\a] \subset [\a+1]$, relevant functions $f$ will lie in many such classes.  One should thus be aware that the definition of the coefficients $f_{pqs}$ depends on the choice of $\a$.}) and similarly for $[\ol{\alpha}]^+$:
\ba
&[\ol\a]^+ [\ol\b] \subset [\ol{\a+\b}]^+,\\
&[\a]^+ [\ol\b] \subset [\a+\b]^+,\\
&[\ol\a]^+ [\ol\b]^+ \subset [\a+\b]^+.
\ea
It is worth noting that $f \in [\a]$ automatically satisfies $f \in [\a]^+$ if either $\a>0$ or $f \in \pa_a [\a-1/2]$.  In other words, we have $[\a]=[\a]^+$ for $\a>0$ (in which case we usually omit the plus sign for simplicity), and $\pa_a [\a-1/2] \subset [\a]$ for any $\a$.

We will now use this terminology to show that the expansions \er{expt}--\er{expg} consistently solve Einstein's equations.  First, from \er{expt}--\er{expg} we find
\be
T \in [\ol 0]^+,\qu
U_i,\, U^i,\, h_{ij},\, h^{ij} \in [\ol 0],
\ee
where condition $\ol{\rm 3}$) is satisfies trivially and indices such as $i$ in $U^i$ are raised using $h^{ij}$, the inverse of the metric $h_{ij}$.  From the metric ansatz \er{mz}, we find the components to satisfy
\ba\la{gzz}
g_{zz} &= T \fr{\zb}{z} \in [\ol 0]^+,\\
g_{z\zb} &= \fr{1}{2} - T \in [\ol 0],\\
g_{zi} &= iU_i \zb \in [\ol{-1/2}],\\
g_{ij} &= h_{ij} \in [\ol 0].\la{gij}
\ea
In establishing \er{gzz}, and in particular that $g_{zz}$ satisfies condition 2) for type $[\ol 0]$, it is important that the expansion \er{expt} for $T$ requires $pq>0$ or $s>0$ and thus includes no purely holomorphic or anti-holomorphic terms.

We similarly find the inverse metric to have components
\ba
g^{zz} &= -\fr{4 T'}{1-4T'} \fr{z}{\zb} \in [\ol 0]^+,\\
g^{z\zb} &= 2+\fr{4T'}{1-4T'} \in 2+[\ol 0]^+ \subset [\ol 0],\\
g^{zi} &= \fr{2iU^i z}{1-4T'} \in [\ol{-1/2}],\\
g^{ij} &= h^{ij}+ \fr{4U^i U^j z\zb}{1-4T'} \in h^{ij} + [\ol{-1}] \subset [\ol 0],
\ea
where $T'\eq T+U^i U_i z\zb \in [\ol 0]^+$, and find the Christoffel symbols $\G_{\r\m\n} = \fr{1}{2} (g_{\r\m,\n}+g_{\r\n,\m}-g_{\m\n,\r})$
to satisfy
\ba
&\G_{abc},\, \T\G{a}{bc} \in [\ol{1/2}],\\
&\G_{abi},\, \G_{iab},\, \T\G{a}{bi},\, \T\G{i}{ab} \in [\ol 0]^+,\\
&\G_{aij},\, \G_{ija},\, \T\G{a}{ij},\, \T\G{i}{ja} \in [\ol{1/2}],\\
&\G_{ijk},\, \T\G{i}{jk} \in [\ol 0].
\ea

The Riemann tensor
\be
R_{\m\n\r\s} = \fr{1}{2} (g_{\m\s,\n\r} + g_{\n\r,\m\s} - g_{\m\r,\n\s} - g_{\n\s,\m\r}) + \G_{\l\m\s} \T\G{\l}{\n\r} - \G_{\l\m\r} \T \G{\l}{\n\s}
\ee
has components
\ba
&R_{z\zb\zb z} \in T_{,z\zb} +\fr{1}{2} \(T \fr{z}{\zb}\)_{,zz} +\fr{1}{2} \(T \fr{\zb}{z}\)_{,\zb\zb} +[1],\\
&R_{z\zb zi} \in \fr{1}{2}\lt\{i(U_i \zb)_{,z\zb} +i(U_i z)_{,zz} -T_{,zi} -\(T \fr{\zb}{z}\)_{,\zb i}\rt\} +[1/2],\\
&R_{z\zb ij} \in [1],\\
&R_{aibj} \in -\fr{1}{2} h_{ij,ab} +[1],\\
&R_{aijk} \in \fr{1}{2} (h_{ij,a;k} - h_{ik,a;j}) +[1/2],\\
&R_{ijkl} \in [1].
\ea
Here indices following commas ($,$), such as the $a$ index in $h_{ij,a;k}$, denote coordinate derivatives and those following semicolons ($;$), such as the $k$ index in $h_{ij,a;k}$, denote covariant derivatives in the $y^k$ direction defined using the $(d-1)$-dimensional metric $h_{ij}$. Here it is important that since $a \in \{z, \zb\}$, the $\partial_a$ operation preserves the tensorial nature of $h_{ij}$ in the $y^k$ directions.  On the other hand, we will use $\na_\mu$ to denote the covariant derivative in the $x^\m$ direction defined using the full spacetime metric $g_{\m\n}$.

The Ricci tensor has components
\ba\la{rzw}
&R_{z\zb} \in 2T_{,z\zb} +\(T \fr{z}{\zb}\)_{,zz} +\(T \fr{\zb}{z}\)_{,\zb\zb} -\fr{1}{2} h^{ij} h_{ij,z\zb} +[1] \subset [\ol 1],\\
&R_{zz} \in -\fr{1}{2} h^{ij} h_{ij,zz} +[1] \subset [\ol 1],\\
&\bmd R_{zi} \in -i(U_i \zb)_{,z\zb} -i(U_i z)_{,zz} +T_{,zi} +\(T \fr{\zb}{z}\)_{,\zb i} +i U^j z h_{ij,zz} -i U^j \zb h_{ij,z\zb}\\
+\fr{1}{2} h^{jk} (h_{ij,z;k} - h_{jk,z;i}) +[1/2] \subset [\ol{1/2}],\emd\\
&R_{ij} \in -2h_{ij,z\zb} +[1] \subset [\ol 1].\la{rij}
\ea

We now solve the vacuum Einstein equations with a cosmological constant\footnote{It is straightforward to generalize the discussion to include matter fields with standard two-derivative actions.} which can be written in the following (trace-reversed) form:
\be\la{emn}
E_{\m\n} \eq R_{\m\n} - \frac{2\Lambda}{d-1} g_{\mu \nu} =0,
\ee
where $\Lambda$ is the cosmological constant.  Inserting \er{gzz}--\er{gij} and \er{rzw}--\er{rij} into \er{emn}, we find that $E_{\m\n}$ satisfies the same equations \er{rzw}--\er{rij} as $R_{\m\n}$.\footnote{For example, $E_{ij}$ differs from $R_{ij}$ only by $\frac{2\Lambda}{d-1} g_{ij} \in [\ol 0] \subset [1]$ which can be absorbed into \er{rij}.}  In other words, we may replace $R$ with $E$ in \er{rzw}--\er{rij}.

First setting $E_{ij}$ to zero at order $z^{\fr{p}{m}} \zb^{\fr{q}{m}} (z\zb)^{s-1}$, we find
\bm\la{rijc}
E_{ij,pqs} = -2\(\fr{p}{m}+s\) \(\fr{q}{m}+s\) h_{ij,pqs} + [1]_{pqs} = 0,\\
\text{for all } p,q,s\geq 0 \text{ with either } pq>0 \text{ or } s>0.
\em
Here $E_{ij,pqs}$ is defined by expanding $E_{ij}$ as a function of type $[\ol 1]$ according to \er{fexp} with $\ell=0$, and for each $i,j$ the object $[1]_{pqs}$ denotes some $f_{pqs}$ defined by the expansion \eqref{fexp} for a function $f$ of type $[1]$.  The values of $f_{pqs}$ for the various $i,j$ need have no relation to each other, and we similarly allow $[1]_{pqs}$ to denote a new coefficient each time it appears below.  

Now, condition 2) for being type $[\ol 1]$ requires that \er{rijc} be trivially satisfied in cases of $pq=s=0$, as can be seen directly from the vanishing of the coefficient $\(\fr{p}{m}+s\) \(\fr{q}{m}+s\)$ and of the second term $[1]_{pqs}$.  In the remaining cases, we use \er{rijc} to solve for $h_{ij,pqs}$ with $pq>0$ or $s>0$ in terms of $T_{p'q's'}$, $U_{i,p'q's'}$, and $h_{ij,p'q's'}$ at lower orders.

Setting $E_{z\zb}$ to zero at order $z^{\fr{p}{m}} \zb^{\fr{q}{m}} (z\zb)^{s-1}$, we find 
\bm\la{rzwc}
E_{z\zb,pqs} = \(\fr{p+q}{m}+2s\) \(\fr{p+q}{m}+2s+1\) T_{pqs} -\fr{1}{2} \(\fr{p}{m}+s\) \(\fr{q}{m}+s\) \T h{ij}{,000} h_{ij,pqs}\\
+ [1]_{pqs} = 0,\qu
\text{for all } p,q,s\geq 0 \text{ with either } pq>0 \text{ or } s>0.
\em
Here $\T h{ij}{,000}$ (as well as the more general $\T h{ij}{,pqs}$ that will appear later is defined by expanding $h^{ij}$ as a function of type $[\ol 0]$ according to \er{fexp} with $\ell=0$.  Again, \er{rzwc} is trivially satisfied in cases of $pq=s=0$.  For $pq>0$ or $s>0$, we may insert  the previously obtained expressions for $h_{ij,pqs}$ and solve for $T_{pqs}$ in terms of $T_{p'q's'}$, $U_{i,p'q's'}$, and $h_{ij,p'q's'}$ at lower orders.

Similarly setting $E_{zi}$ to zero at order $z^{\fr{p}{m}-1} \zb^{\fr{q}{m}} (z\zb)^s$, we find 
\bm\la{rzic}
E_{zi,pqs} = -i \(\fr{p}{m}+s\) \(\fr{p+q}{m}+2s+2\) U_{i,pqs} +\(\fr{p+q}{m}+2s+1\) T_{pqs,i}\\
+i \(\fr{p}{m}+s\) \(\fr{p-q}{m}-1\) U_{j,000} \T h{jk}{,000} h_{ik,pqs} +\fr{1}{2} \(\fr{p}{m}+s\) \T h{jk}{,000} (h_{ij,pqs;k}-h_{jk,pqs;i})\\
+ [1/2]_{pqs} =0,\qu
\text{for all } p,q,s\geq 0 \text{ with either } p>0 \text{ or } s>0.
\em
Here, in a slight change of notation, indices after a semicolon ($;$) denote covariant derivatives defined using the metric $h_{ij,000}$. We will use this definition to take covariant derivatives of individual coefficients with subscript $pqs$ in an expansion of the form \eqref{fexp} as opposed to taking covariant derivatives of the full sum.  Note that $E_{zi}$ is of type $[\ol{1/2}]$ with $\ell=-1$, so \er{rzic} is trivially satisfied in cases of $p=s=0$.  For $p>0$ or $s>0$, we may insert the previously obtained expressions for $T_{pqs}$, $h_{ij,pqs}$ and solve for $U_{i,pqs}$ in terms of $T_{p'q's'}$, $U_{i,p'q's'}$, and $h_{ij,p'q's'}$ at lower orders.

Setting $E_{\zb i}$ to zero at order $z^{\fr{p}{m}} \zb^{\fr{q}{m}-1} (z\zb)^s$, we find the complex conjugate of \er{rzic} with $p$ and $q$ exchanged in the coefficients:
\bm\la{rwic}
E_{\zb i,pqs} = i \(\fr{q}{m}+s\) \(\fr{p+q}{m}+2s+2\) U_{i,pqs} +\(\fr{p+q}{m}+2s+1\) T_{pqs,i}\\
-i \(\fr{q}{m}+s\) \(\fr{q-p}{m}-1\) U_{j,000} \T h{jk}{,000} h_{ik,pqs} +\fr{1}{2} \(\fr{q}{m}+s\) \T h{jk}{,000} (h_{ij,pqs;k}-h_{jk,pqs;i})\\
+[1/2]_{pqs} =0,\qu
\text{for all } p,q,s\geq 0 \text{ with either } q>0 \text{ or } s>0.
\em
This equation is trivially satisfied in cases of $q=s=0$.  For $q>0$ or $s>0$, we may insert the previously obtained expressions for $T_{pqs}$, $h_{ij,pqs}$ and solve for $U_{i,pqs}$ in terms of $T_{p'q's'}$, $U_{i,p'q's'}$, and $h_{ij,p'q's'}$ at lower orders.  However, some of these $U_{i,pqs}$ (those with $pq>0$ or $s>0$) have already been determined from \er{rzic}, and we will need to show that the two solutions agree.  This is true and guaranteed by the contracted Bianchi identities, as we will show in a moment.

Setting $E_{zz}$ to zero at order $z^{\fr{p}{m}-2} \zb^{\fr{q}{m}} (z\zb)^{s}$, we find
\bm\la{rzzc}
E_{zz,pqs} = -\fr{1}{2} \(\fr{p}{m}+s\) \(\fr{p}{m}+s-1\) \T h{ij}{,000} h_{ij,pqs} + [1]_{pqs} = 0,\\
\text{for all } p,q,s\geq 0 \text{ with either } p>0 \text{ or } s>1.
\em
Here $E_{zz}$ is of type $[\ol 1]$ with $\ell=-2$, so \er{rzzc} is trivially satisfied in cases with $p=0$, $s\leq 1$.  In cases with $p>0$ and $q=s=0$, we use \er{rzzc} to express the trace $\T h{ij}{,000} h_{ij,p00}$ in terms of $T_{p'q's'}$, $U_{i,p'q's'}$, and $h_{ij,p'q's'}$ at lower orders.  In the remaining cases, \er{rzzc} is guaranteed by equations that we have already satisfied, as we will show in a moment using the contracted Bianchi identities.

Finally setting $E_{\zb\zb}$ to zero at order $z^{\fr{p}{m}} \zb^{\fr{q}{m}-2} (z\zb)^{s}$, we find the complex conjugate of \er{rzzc} with $p$ and $q$ exchanged in the coefficients:
\bm\la{rwwc}
E_{\zb\zb,pqs} = -\fr{1}{2} \(\fr{q}{m}+s\) \(\fr{q}{m}+s-1\) \T h{ij}{,000} h_{ij,pqs} + [1]_{pqs} = 0,\\
\text{for all } p,q,s\geq 0 \text{ with either } q>0 \text{ or } s>1.
\em
This equation is trivially satisfied in cases of $q=0$, $s\leq 1$.  Again, in cases with $q>0$ and $p=s=0$ we use \er{rwwc} to write the trace $\T h{ij}{,000} h_{ij,0q0}$ in terms of $T_{p'q's'}$, $U_{i,p'q's'}$, and $h_{ij,p'q's'}$ at lower orders.  In the remaining cases, \er{rwwc} will be guaranteed by the contracted Bianchi identities.

We now turn to the  contracted Bianchi identities
\be\la{cbi}
B_\m \eq \na_\n \T R{\n}{\m} - \fr{1}{2} \na_\m R = \na_\n \T E{\n}{\m} - \fr{1}{2} \na_\m E =0
\ee
and, given our earlier solutions, show that for $\m=i$ they guarantee $E_{\zb i,pqs}=0$ with $pq>0$ or $s>0$, for $\m=z$ they guarantee $E_{zz,pqs}=0$ with $q>0$ or $s>0$, and for $\m=\zb$ they guarantee $E_{\zb\zb,pqs}=0$ with $p>0$ or $s>0$ as claimed above.  To see this, note that \er{cbi} leads to
\ba\label{Beqs}
B_{i,pqs} &= 2 \(\fr{p}{m}+s\) E_{\zb i,pqs} + \cd =0,\\
B_{z,pqs} &= 2 \(\fr{q}{m}+s\) E_{zz,pqs} + \cd =0,\\
B_{\zb,pqs} &= 2 \(\fr{p}{m}+s\) E_{\zb\zb,pqs} + \cd =0,
\ea
where $\cd$ denotes terms that are linear combinations of $E_{\zb i,p'q's'}$, $E_{zz,p'q's'}$, and $E_{\zb\zb,p'q's'}$ at lower orders, as well as $E_{ij,p'q's'}$, $E_{zi,p'q's'}$, and $E_{z\zb,p'q's'}$ of any orders.  The desired conclusion then follows immediately.

In summary, we have now shown that the expansions \er{expt}--\er{expg} consistently solve Einstein's equations, at least in the generic case where $m$ is irrational.  Before discussing rational $m$, note that it is possible to find an exact expression for $E_{zz,p00}$ that may then be used to solve for the traces $\T h{ij}{,000} h_{ij,p00}$:
\bm\la{rzzp}
E_{zz,p00} = R_{zz,p00} =  -\sum_{p_1=1}^{p} \fr{p_1}{2m} \(\fr{p_1}{m}-1\) h_{ij,p_1 00} \T h{ij}{,(p-p_1)00}\\
+\sum_{\substack{p_1,p_2>0,\, p_3\geq 0\\p_1+p_2+p_3\leq p}} \fr{p_1 p_2}{4 m^2} h_{ij,p_1 00} h_{kl,p_2 00} \T h{ik}{,p_3 00} \T h{jl}{,(p-p_1-p_2-p_3)00} = 0,\qu
\forall\, p>0.
\em
Similarly for $E_{\zb\zb,0q0}$ we have
\bm\la{rwwq}
E_{\zb\zb,0q0} = R_{\zb\zb,0q0} = -\sum_{q_1=1}^{q} \fr{q_1}{2m} \(\fr{q_1}{m}-1\) h_{ij,0 q_1 0} \T h{ij}{,0(q-q_1)0}\\
+\sum_{\substack{q_1,q_2>0,\, q_3\geq 0\\q_1+q_2+q_3\leq q}} \fr{q_1 q_2}{4 m^2} h_{ij,0 q_1 0} h_{kl,0 q_2 0} \T h{ik}{,0 q_3 0} \T h{jl}{,0(q-q_1-q_2-3_3)0} = 0,\qu
\forall\, q>0.
\em
These two equations simplify for $p=1$ and $q=1$ respectively to yield
\ba
E_{zz,100} &= -\fr{1}{2m} \(\fr{1}{m}-1\) \T h{ij}{,000} h_{ij,100} = 0\qu
\Rightarrow\qu
\T h{ij}{,000} h_{ij,100} = 0,\\
E_{\zb\zb,010} &= -\fr{1}{2m} \(\fr{1}{m}-1\) \T h{ij}{,000} h_{ij,010} = 0\qu
\Rightarrow\qu
\T h{ij}{,000} h_{ij,010} = 0.
\ea
Defining $K_{zij} \eq \frac{1}{2}h_{ij,100}$ and $K_z \eq \T h{ij}{,000} K_{zij}$, and making the corresponding definitions for complex conjugates, we find
\be\la{kzz}
K_z = K_\zb = 0.
\ee
This condition is valid for general values of $m$, and taking the limit $m\to 1$ it becomes the familiar requirement that the trace of the extrinsic curvature tensor vanish.

Now, even with a conical singularity, in Euclidean signature it is a well-defined question to ask whether the area of a given surface is locally minimal with respect to small deformations.  Here it is important to realize that, even in smooth Euclidean spacetimes, extremal surfaces are not necessarily locally minimal in this sense\footnote{Such examples are directly analogous to a geodesic in 2-dimensional space over the top of a hill.}, but can instead give more general saddles.  But all extremal surfaces in smooth Euclidean geometries {\it are} locally minimal with respect to variations that are also sufficiently local in space -- i.e., where most of the surface is held fixed and only an arbitrarily small piece of the surface is allowed to vary. We will refer to variations of this sort as doubly-local.

It is thus of interest to ask how the condition \eqref{kzz} relates to the possibility that the conical singularity may lie on a surface of doubly-locally minimal area.
Let us consider such a doubly-local variation of a surface from the conical defect to a nearby location parameterized by $z=\e \td z(y^i)$.
From the metric \er{mz} and the expansions \eqref{expt}--\eqref{expg}, we find that without imposing Einstein's equations the area generally changes by $O(\e^2)$ and $O(\e^{1/m})$ effects.
The coefficient of the $O(\epsilon^2)$ term is positive when the variation is sufficiently localized in the directions along the surface. For $m \le \frac{1}{2}$, the leading area change is $O(\e^2)$ and doubly-local minimality follows directly as in smooth Euclidean geometries\footnote{Although we have not yet discussed rational $m$,  for the case of integer $n=1/m$, this is clear from the fact that the $n$-fold cover is a smooth geometry with a replica ${\mathds Z}_n$ symmetry about the would-be singularity.  This symmetry then requires the would-be singularity to be extremal, and thus to be doubly-locally minimal. Indeed, the full extrinsic curvature tensor must vanish by symmetry. It follows that the conical singularity lies on a doubly-locally minimal surface in the ${\mathds Z}_n$ quotient.  It is worth noting that \er{kzz} in this case imposes an additional trace condition for not the extrinsic curvature but a higher-order version of it (i.e., $\pa_z^n h_{ij}$ and its complex conjugate).\label{footm}}.
For $\frac{1}{2} < m \leq 1$, doubly-locally minimality would have failed without the equations of motion, but once we impose them, \er{kzz} forces the $O(\e^{1/m})$  terms in the area change to vanish, ensuring that the leading area change is still $O(\e^2)$ and doubly-local minimality holds.  For $m>1$, the leading area change is generally dominated by $O(\e^{2/m})$ effects involving quadratic terms in $h_{ij,100}$ and $h_{ij,010}$ which do not have a definite sign, so the conical defect is not doubly-locally minimal in this case.  Nonetheless, \er{kzz} holds in this case (as in the previous two cases), and it imposes a nontrivial constraint on on-shell geometries that postpones a potential $O(\e^{1/m})$ change in the area to $O(\e^{2/m})$.

Let us now consider the special case where $m$ is a rational number.  We could work out this case directly by following a procedure similar to that above, but it is easier to take a limit where $m$ approaches a rational number from a sequence of irrational numbers.  To see that the limit is well-behaved, note that when we solve the equations of motion \er{rijc}--\er{rwwc}, the coefficients of the terms for which we solve are continuous functions of $m$ that do not generally vanish at rational $m$.  Indeed, the only exception is the coefficient $-\fr{1}{2} \(\fr{p}{m}+s\) \(\fr{p}{m}+s-1\)$ in \er{rzzc}, which vanishes for $q=s=0$ when $m$ approaches the positive integer $p$.  In this case, \er{rzzc} is given by the more precise version \er{rzzp}, and instead of solving for the trace $\T h{ij}{,000} h_{ij,p00}$ we will simply leave the trace undetermined and interpret \er{rzzp} with $p=m$ as a constraint on the coefficients $h_{ij,p'00}$ with $p'<m$ for which we did not solve above. For $p=m>1$, this constraint is manifestly nontrivial\footnote{This does not apply to $p=m=1$ for which \er{rzzp} is trivially satisfied and leads to no constraint.  In this case, our problem reduces to finding standard smooth solutions, and the ``missing'' constraint from \er{rzzp} -- as well as its complex conjugate -- is explained by the additional diffeomorphism gauge invariance associated with moving the location of the codimension-2 surface marked by $r=0$ (which is possible only when there is no conical defect).} as can be seen from \er{rzzp}, and a solution to the constraint would typically exist.  As a result, the total number of free parameters will be the same as in more generic cases of non-integer $m$.  This is the key point that we require for the discussion of uniqueness of solutions in section \ref{subsec:unique} below.

\sse{Uniqueness of solutions}
\label{subsec:unique}

We now give a counting argument to show that the freedom in the solutions constructed in the previous subsection is precisely what one expects to need to match general boundary conditions at large $r$.  In other words, we will show that there are enough solutions of the form \eqref{expt}--\er{expg} to describe the expected physics, and also that solutions of this form that are compatible with given large-$r$ boundary conditions generally have no continuous free parameters.  Instead, such solutions form a discrete set as one expects of a good non-linear elliptic boundary value problem.

In the Asymptotically locally AdS (AlAdS) context, one generally requires the induced geometry on a constant $r$ slice to be conformal to a given $d$-dimensional boundary metric in the limit $r\to\infty$.  The boundary metric has $d(d+1)/2$ independent components, involving $d(d+1)/2$ general functions of $(\p,y^i)$.  However, when matching to the boundary metric we can use any conformal factor\footnote{Alternatively, one can fix the conformal factor but allow the freedom to use surfaces $\Sigma_\epsilon$ defined by $r=\bar r(\p,y^i)/\e$ which nevertheless approach $r=\infty$ as $\epsilon \rightarrow 0$, with the same counting due to the arbitrary function $\bar r(\p,y^i)$.  This latter formulation is preferred in odd bulk dimensions due to the boundary conformal anomaly \cite{Papadimitriou:2005ii}.  Similarly, one might instead impose a finite-distance Dirichlet boundary condition, requiring that there be a surface at finite distance with a fixed induced geometry, though in that case the coordinate location of the surface should not be fixed.} and any $d$-dimensional diffeomorphism to identify the constant $r$ slice with the given conformal geometry, and these are parameterized by $d+1$ general functions of $(\p,y^i)$.  Therefore, the asymptotic boundary conditions are parameterized by $(d+1)(d-2)/2$ functions of $(\p,y^i)$.

We now show that, after removing residual gauge transformations, the free parameters in the small $r$ expansions \er{expt}--\er{expg} are also precisely $(d+1)(d-2)/2$ functions of $(\p,y^i)$.  To begin, note that at non-integer $m$ our procedure for solving the equations of motion \er{rijc}--\er{rwwc} expressed the solution in terms of the following unconstrained coefficients:
\begin{enumerate}
\item $U_{i,000}$: We refer to these coefficients as $d-1$ ``zero modes,'' by which we mean that they are functions of the $y^i$ alone and have vanishing angular momentum on the $\phi$ circle.
\item $h_{ij,000}$: These give an additional $d(d-1)/2$ zero modes.
\item The traceless parts of $h_{ij,p00}$, $h_{ij,0q0}$ for any $p, q > 0$: Since a general periodic function of $\phi$ can be expanded in a Fourier series, these coefficients can be equivalently expressed as $d(d-1)/2-1$ functions of $(\p,y^i)$ whose components at zero angular momentum on the $\phi$ circle are constrained to vanish; i.e., they are missing the corresponding zero modes.
\end{enumerate}
Putting these together, the remaining free data consists of $d(d-1)/2-1 = (d+1)(d-2)/2$ functions of $(\p,y^i)$ (now with freely specifiable zero modes), together with $d$ additional zero modes.

To proceed, we must also count residual gauge transformations.  These are diffeomorphisms that preserve the form of the metric ansatz \er{mz}.  They consist of $(d-1)$-dimensional diffeomorphisms in the $y^i$ directions, as well as arbitrary $y^i$-dependent shifts of the $\p$ coordinate: $\p \to \p+ \x(y^i)$.  In total, these residual gauge transformations are parameterized by $d$ zero modes, which we should subtract from the number of free parameters in the small $r$ expansions (because the residual gauge transformations preserve the asymptotic boundary conditions up to conformal factors and boundary diffeomorphisms).

Up to residual gauge transformations, the free parameters in the small $r$ expansions can thus be expressed as $(d+1)(d-2)/2$ functions of $(\p,y^i)$.  This precisely matches the freedom in the large-$r$ boundary conditions.  For non-integer $m$,  the solution constructed in the previous subsection thus contains precisely the right amount of freedom to solve the desired boundary value problem.  Indeed, modulo residual gauge transformations, for given such boundary conditions the solutions will generally admit no continuous parameters, and will thus form a discrete set as expected of a good elliptic boundary-value problem.

The special case of integer $m$ is much the same.  As noted at the end of section \ref{subsec:gen}, in that case our procedure leaves the trace $\T h{ij}{,000} h_{ij,p00}$ undetermined, and instead enforces a different constraint on the coefficients $h_{ij,p'00}$ with $p'<m$.  Although this case does not organize itself as nicely into the Fourier transform of $d(d-1)/2-1 = (d+1)(d-2)/2$ functions, it contains the same number of free parameters.  Furthermore, since we are attempting to match to boundary conditions at large $r$, and since the equations of motion are non-linear, the free parameters we find should generally be expected to match all Fourier components of the boundary data as desired.  A counting argument of this form is thus the best one can expect to achieve at this level of analysis.  Even in the smooth case $m=1$, to our knowledge there is no theorem guaranteeing the existence of solutions with arbitrary boundary data.

\section{Higher derivative variational principles by minimal subtraction}
\label{app:HDI}

In this appendix, we generalize the discussion in appendix \re{app:LMVPRT} for Einstein gravity to include arbitrary higher-derivative corrections.  The action is defined in a similar way as in \er{ig}:
\be\la{igh}
\td I[g] = \lim_{\e\to0^+} \lt\{\int_{r\geq\e} d^{d+1}x \sqrt{g} \cL + I_{\text{CT}}^\e\rt\},
\ee
where the higher-derivative Lagrangian has the general form
\be\la{lh}
\cL = -\fr{1}{8\pi G}\(\fr{R-2\L}{2} + \l_1 R^2 +\l_2 R_{\m\n} R^{\m\n} +\l_3 R_{\m\n\r\s} R^{\m\n\r\s} +\l_4 R\na^2 R +\cd\)
\ee
and $I_{\text{CT}}^\e$ is an appropriate counterterm to be specified later.

We will work in the perturbative limit where higher-derivative corrections are small and physical quantities can be solved as Taylor expansions in the higher-derivative coupling constants $\l_k$.  In particular, the metric has the form
\be
g_{\m\n} = \sum_{n_1, n_2, \cd =0}^\infty g_{\m\n}^{{(n_1 n_2\cd)}} \l_1^{n_1} \l_2^{n_2} \cd.
\ee
We we will sometimes abbreviate $g_{\m\n}^{{(n_1 n_2\cd)}}$ as $g_{\m\n}^{(\vn)}$.

As before, we may choose quasi-cylindrical coordinates so that the metric is of the form \er{mz} to any order in the perturbative expansion.  However, the corresponding functions $T^{(\vn)}$, $U_i^{(\vn)}$, and $h_{ij}^{(\vn)}$ generally have more singular behaviors at $r=0$ than indicated in the expansions \er{expt}--\er{expg}.  To see this precisely, let us start again with the generic case where $m$ is an irrational number.  Instead of the expansions \er{expt}--\er{expg}, we will show
\ba\la{exptn}
T^{(\vn)} &= \sum_{\substack{p,q,s=0\\pq>0 \text{ or } s>n}}^{\infty} T^{(\vn)}_{pqs} z^{\fr{p}{m}} \zb^{\fr{q}{m}} (z\zb)^{s-n},\\
U_i^{(\vn)} &= \sum_{\substack{p,q,s=0\\pq>0 \text{ or } s\geq n}}^{\infty} U^{(\vn)}_{i,pqs} z^{\fr{p}{m}} \zb^{\fr{q}{m}} (z\zb)^{s-n},\\
\la{expgn}
h_{ij}^{(\vn)} &= \sum_{\substack{p,q,s=0\\pq>0 \text{ or } s\geq n}}^{\infty} h^{(\vn)}_{ij,pqs} z^{\fr{p}{m}} \zb^{\fr{q}{m}} (z\zb)^{s-n}.
\ea
Here $n$ is a nonnegative number determined by $\vn$:
\be
n = \sum_{k=1}^\infty n_k \(\fr{D_k}{2}-1\),
\ee
where $D_k$ is the total number of derivatives in the term whose coefficient in the Lagrangian \er{lh} is $\l_k$.  For example, we have $D_1=D_2=D_3=4$ for the $4$-derivative terms and $D_4=6$ for the $6$-derivative term in \er{lh}.

We say that a function $f$ is of type $[\a]^{(\vn)}$ if it satisfies conditions 1) and 2) used previously in appendix \re{app:LMVPRT} to define type $[\a]$, but instead of 3) it satisfies the following generalization:
\begin{enumerate}
\item[3${}^{\vn}$)] Each coefficient $f_{pqs}$ only depends on $T^{(\vn')}_{p'q's'}$, $U^{(\vn')}_{i,p'q's'}$, and $h^{(\vn')}_{ij,p'q's'}$ at lower orders, meaning either $\vn'=\vn$ and $(p',q',s') < (p,q,s)$ as defined in \eqref{<def}, or $\vn' < \vn$.  Here $\vn' < \vn$ is defined by the conditions
\be
\qu n'_k \leq n_k \, \text{ for all } \, k \geq 1\qu
\text{and}\qu \vn'\neq \vn.
\ee
\end{enumerate}
We define type $[\ol\a]^{(\vn)}$ in the same way as $[\a]^{(\vn)}$, except that in condition 3${}^{\vn}$) we allow lower or equal orders, meaning either $\vn'=\vn$ and $(p',q',s') \leq (p,q,s)$ as defined in \er{pqsle}, or $\vn' < \vn$.

Using this terminology and working at any perturbative order, we find the metric components
\ba
g^{(\vn)}_{zz} &= T^{(\vn)} \fr{\zb}{z} \in [\ol n]^{(\vn)},\\
g^{(\vn)}_{z\zb} &= \fr{1}{2}\d_{\vn,0} - T^{(\vn)} \in [\ol n]^{(\vn)},\\
g^{(\vn)}_{zi} &= iU^{(\vn)}_i \zb \in [\ol{n-1/2}]^{(\vn)},\\
g^{(\vn)}_{ij} &= h^{(\vn)}_{ij} \in [\ol n]^{(\vn)},
\ea
the Riemann tensor components
\ba
&R^{(\vn)}_{z\zb\zb z} \in T^{(\vn)}_{,z\zb} +\fr{1}{2} \(T^{(\vn)} \fr{z}{\zb}\)_{,zz} +\fr{1}{2} \(T^{(\vn)} \fr{\zb}{z}\)_{,\zb\zb} +[n+1]^{(\vn)},\\
&R^{(\vn)}_{z\zb zi} \in \fr{1}{2}\lt\{i\(U^{(\vn)}_i \zb\)_{,z\zb} +i\(U^{(\vn)}_i z\)_{,zz} -T^{(\vn)}_{,zi} -\(T^{(\vn)} \fr{\zb}{z}\)_{,\zb i}\rt\} +[n+1/2]^{(\vn)},\\
&R^{(\vn)}_{z\zb ij} \in [n+1]^{(\vn)},\\
&R^{(\vn)}_{aibj} \in -\fr{1}{2} h^{(\vn)}_{ij,ab} +[n+1]^{(\vn)},\\
&R^{(\vn)}_{aijk} \in \fr{1}{2} \(h^{(\vn)}_{ij,a;k} - h^{(\vn)}_{ik,a;j}\) +[n+1/2]^{(\vn)},\\
&R^{(\vn)}_{ijkl} \in [n+1]^{(\vn)},
\ea
and the Ricci tensor components
\ba
&R^{(\vn)}_{z\zb} \in 2T^{(\vn)}_{,z\zb} +\(T^{(\vn)} \fr{z}{\zb}\)_{\!\!,zz} +\(T^{(\vn)} \fr{\zb}{z}\)_{\!\!,\zb\zb} -\fr{1}{2} h^{(0)ij} h^{(\vn)}_{ij,z\zb} +[n+1]^{(\vn)} \subset [\ol{n+1}]^{(\vn)},\\
&R^{(\vn)}_{zz} \in -\fr{1}{2} h^{(0)ij} h^{(\vn)}_{ij,zz} +[n+1]^{(\vn)} \subset [\ol{n+1}]^{(\vn)},\\
&\bmd R^{(\vn)}_{zi} \in -i\(U^{(\vn)}_i \zb\)_{,z\zb} -i\(U^{(\vn)}_i z\)_{,zz} +T^{(\vn)}_{,zi} +\(T^{(\vn)} \fr{\zb}{z}\)_{,\zb i} +i U^{(0)j} z h^{(\vn)}_{ij,zz}\\
-i U^{(0)j} \zb h^{(\vn)}_{ij,z\zb} +\fr{1}{2} h^{(0)jk} \(h^{(\vn)}_{ij,z;k} - h^{(\vn)}_{jk,z;i}\) +[n+1/2]^{(\vn)} \subset [\ol{n+1/2}]^{(\vn)},\emd\\
&R^{(\vn)}_{ij} \in -2h^{(\vn)}_{ij,z\zb} +[n+1]^{(\vn)} \subset [\ol{n+1}]^{(\vn)}.
\ea
Here covariant derivatives in the $y^k$ directions in terms such as $h^{(\vn)}_{ij,a;k}$ are defined using the $(d-1)$-dimensional metric $h^{(0)}_{ij}$ at the zeroth order in higher-derivative coupling constants.

One immediate consequence is that for any covariant scalar $f$ built from a product of an arbitrary number of the metric $g_{\m\n}$, the inverse metric $g^{\m\n}$, and the Riemann tensor $R_{\m\n\r\s}$ with possible covariant derivatives, we have
\be\la{fn}
f^{(\vn)} \in [\ol{n+D/2}]^{(\vn)}
\ee
where $D$ is the total number of derivatives in the expression $f$.

The equation of motion including higher-derivative interactions can be written in the following (trace-reversed) form:
\be\la{eh}
E_{\m\n} \eq R_{\m\n} -\fr{2\L}{d-1} g_{\m\n} +\sum_{k=1}^\infty \l_k E_{(k)\m\n} =0,
\ee
where $E_{(k)\m\n}$ is the contribution from the term with coefficient $\l_k$ in the Lagrangian \er{lh}, and is a sum of terms each with no more than $D_k$.  Using the same strategy as in appendix \re{app:LMVPRT}, we find that the expansions \er{exptn}--\er{expgn} can consistently solve the equation of motion at any perturbative order.  We will not repeat all the details here, but as an example we find from \er{eh} that $E^{(\vn)}_{ij}$ is of type $[\ol{n+1}]^{(\vn)}$, and upon setting it to zero at order $z^{\fr{p}{m}} \zb^{\fr{q}{m}} (z\zb)^{s-n-1}$ we get
\bm\la{eijc}
E^{(\vn)}_{ij,pqs} = -2\(\fr{p}{m}+s-n\) \(\fr{q}{m}+s-n\) h^{(\vn)}_{ij,pqs} +[n+1]^{(\vn)}_{pqs} = 0,\\
\forall\, p,q,s\geq 0, \text{ either } pq>0 \text{ or } s>n.
\em
Condition 2) of $E^{(\vn)}_{ij}$ being type $[\ol{n+1}]^{(\vn)}$ requires that \er{eijc} be trivially satisfied in cases of $pq=0$ and $s\leq n$, as can be seen directly from the vanishing of both terms in \er{eijc}.  The vanishing of the first term is due to the vanishing of either its coefficient $\(\fr{p}{m}+s-n\) \(\fr{q}{m}+s-n\)$ for $s=n$ or $h^{(\vn)}_{ij,pqs}$ for $s<n$ as is clear from its definition \er{expgn}.  In the remaining cases, we use \er{eijc} to solve for $h^{(\vn)}_{ij,pqs}$ with $pq>0$ or $s>n$ in terms of $T^{(\vn')}_{p'q's'}$, $U^{(\vn')}_{i,p'q's'}$, and $h^{(\vn')}_{ij,p'q's'}$ at lower orders.

The other components of the equation of motion can be solved similarly.  Using the same counting argument as in appendix \re{app:LMVPRT}, we find that at any order in the higher-derivative coupling constants, the solution constructed here is generically unique up to residual gauge transformations once we impose suitable asymptotic boundary conditions.

We now show (still for irrational $m$) that the higher-derivative action \er{igh} again leads to a well-defined variational principle for metric configurations of the form \er{mz} with a fixed opening angle $2\pi m$ on the conical defect, once we choose the counterterm $I_{\text{CT}}^\e$ appropriately.  This works at any order in the higher-derivative coupling constants, and the metric configurations follow the expansions \er{exptn}--\er{expgn} at $r=0$.  For such metric configurations, we find from \er{fn} that the Lagrangian \er{lh} generally has singular behaviors at $r=0$ characterized by
\be
\cL^{(\vn)} \in [\ol{n+1}]^{(\vn)}
\ee
whose integral is generally divergent at $r=0$ for $n>0$.  However, condition 2) of being type $[\ol{n+1}]^{(\vn)}$ ensures that $\cL^{(\vn)}$ does not have negative integer powers of $z$ or $\zb$, at least in the generic case where $m$ is irrational.  Therefore, the integral in \er{igh} has only power-law (but not logarithmic) divergences.  In particular, the divergences at the $\vn$th perturbative order are of the form $\e^{2\(\fr{p}{m}-s\)}$ where $s \leq n$, $p$ are positive integers.  We may thus choose the counterterm $I_{\text{CT}}^\e$ so that it minimally subtracts these power-law divergences, yielding a finite action \er{igh}.

With this choice of the counterterm, we find a well-defined variational principle under the boundary condition that fixes $m$ to any irrational value.  To see this, note that under a general, infinitesimal variation $\d g_{\m\n}$ of the metric, the action $\er{igh}$ changes by a boundary term at $r=\e$:
\be\la{digh}
\d \td I[g] = \lim_{\e\to0^+} \[ \lt. \int_\pa d^dX \sqrt{\g} n^{\m} V_\m \rt|_{r=\e} + \d I_{\text{CT}}^\e \],
\ee
up to a bulk term that vanishes if the equation of motion is satisfied.  The notation here is similar to what was used in \er{dig}, and $V_\m$ is a vector built from $g_{\m\n}$, $g^{\m\n}$, $\d g_{\m\n}$, $R_{\m\n\r\s}$, and their covariant derivatives.  
At least in the generic case of irrational $m$, terms with negative integer powers of $z$ or $\zb$ cannot appear in metric variations $\d g_{\m\n}$ that fix $m$; hence they also cannot appear in $V_\m$.  At any perturbative order, any $\e\to 0$  divergences in the first term on the right hand side of \er{digh} are thus power laws which must be precisely cancelled by $\d I_{\text{CT}}^\e$; after all, $\d \td I[g]$ cannot be infinite if $\td I[g]$ is finite.  The important point is that the integral in \er{digh} cannot have a finite, nonzero term as $\e\to 0$.  To see this, note that $\sqrt{\g}$ is $r$ times an expression that is built from $T$, $U_i$, and $h_{ij}$ and therefore has no negative integer powers of $z$ or $\zb$ according to condition 2), $V_\m$ similarly has no negative integer powers of $z$ or $\zb$, and the unit normal vector is specified by $(n^z, n^\zb, n^i)=(z, \zb, 0)/r$.  Therefore, the boundary term \er{digh} vanishes, leading to a well-defined variational principle for fixed irrational $m$.

As before, we note that varying the on-shell action with respect to $m$ must give some geometric invariant integrated on the conical defect.   For any (irrational) value of $m$ this invariant is a higher-derivative generalization of the area.  This arises from the boundary term \er{digh}, but is different from the fixed-$m$ variations discussed above because changing $m$ in \eqref{met} introduces a nonzero $T_{000}$, leading to a $\zb/z$ term in the $zz$ component of the metric (as well as its complex conjugate).  It also introduces $\log z, \log \zb$ terms, but since $\tilde I[g]$ is finite at each $m$ these must either cancel in \eqref{digh} or vanish as $\epsilon \rightarrow 0$. For our purposes we do not need to work out the explicit form of this boundary term.  Instead, we simply define it to be $-\s \d m$, leading to
\be\la{dim}
\fr{d \td I_m}{d m} = -\s,
\ee
where $\td I_m$ again denotes the on-shell action with the boundary condition set by $m$.  Note also that the form of \eqref{digh} requires \eqref{dim} to be given by a boundary term at the defect independent of both $z$ and $\zb$, so that (if desired) $\sigma$ may be computed by taking the limit $\epsilon \rightarrow 0$ from any fixed direction in the bulk spacetime.

We expect the $zz$ and $\zb\zb$ components of the equation of motion to lead to higher-derivative generalizations of the vanishing trace of the extrinsic curvature tensor in \er{kzz}.  In particular, taking the limit $m\to1$ we expect to find that the HRT surface extremizes $\s$.  Instead of working out these details by brute force, we note that we can use the above variational principle to solidify the argument outlined in \cite{Dong:2017xht} and in footnote \ref{foot3}.  To do so, we first consider a metric variation $\d g_{\m\n}$ that for $m=1$ corresponds to an infinitesimal but arbitrary change of the location of the HRT surface.  Note that, consistent with \eqref{met}, we will {\it not} change coordinates but will instead change the induced metric $h_{ij}$ as well as $U_i$ and $T$ in the manner defined by a diffeomorphism that acts non-trivially on the HRT surface.  Note that this change in $(h_{ij},U_i, T)$ also defines a valid variation $\delta g_{\m\n}$ for $m=1$, though the latter need not always be equivalent to acting with a diffeomorphism.  We can then apply a second infinitesimal variation that changes $m$ from $1$ to $1+d m$.  But the two variations $d$ and $\d$ commute.  And according to \er{dim}, when acting on $\tilde I_m$ the former variation gives $-\sigma$.  We thus find
\be
\fr{d \d \td I_m}{d m}  = \d \fr{d  \td I_m}{d m}  = -\d\s.
\ee
Furthermore, 
the left-most expression must vanish as for any $m$ the quantity $\d \td I_m$ vanishes under all variations $\d g_{\m\n}$ that fix $m$.  From this we find $\d\s=0$ under an arbitrary shift of the HRT surface, so $\sigma$ is extremized on-shell at $m=1$. A similar argument shows that in limits $m\rightarrow 1/n$ for integer $n$, in which case the limit has a smooth $n$-fold cover, the geometric entropy $\sigma$ is extremized in the covering space.  

In parallel with the discussion in the two paragraphs below \eqref{kzz}, one might also ask whether the conical singularity in our solutions also sits on a surface that double-locally minimizes $\sigma$ (i.e., it minimizes $\sigma$ with respect to variations that are localized in directions along the surface as well as transverse to the surface).  At a very formal level the results would seem to be the same as for our previous discussion of the area in conical spacetimes (and for the case of $n=1/m$, doubly-local minimality again follows directly by symmetry as in footnote \re{footm}).  But to give a precise argument for general $m$, one would need to think carefully about how to define $\sigma$ for smooth surfaces that intersect the conical singularity.  We leave this issue for future investigation.

Finally, let us comment on the special case where $m$ is a rational number.  As before, we take a limit where $m$ approaches a rational number from a sequence of irrational numbers.  However, the limit here is not necessarily well-behaved, because when solving various components of the equation of motion such as \er{eijc}, the coefficients of the terms for which we solve involve expressions like $\(\fr{p}{m}+s-n\) \(\fr{q}{m}+s-n\)$ which may vanish as $m$ approaches a rational number.  In Einstein gravity we solved a similar problem by requiring the rest of the equation to vanish, but here we do not generally have that freedom because the rest of the equation is sometimes determined completely by solutions at lower orders in the higher-derivative coupling constants.  This means that at rational values of $m$, the perturbative expansion in the higher-derivative coupling constants may develop a pole in $m$ at some order $\vec n$.  However, this breakdown of perturbation theory never happens when $m$ is the inverse of a positive integer -- since the solution in that case can be constructed as the $\bZ_{1/m}$ quotient of a smooth geometry.  Moreover, at any given perturbative order -- for example the $\vn$th order -- this breakdown occurs on at most a nowhere dense set of rational values of $m$, with a minimal distance set by $1/n$.

\bibliographystyle{JHEP}
\bibliography{Renyi}

\end{document}